\documentclass[prb,superscriptaddress,showpacs,floatfix,onecolumn]{revtex4}
\usepackage{amssymb}

\usepackage{amsmath}
\usepackage{graphicx}

\input{tcilatex}

\begin{document}

\title{Commensurate-incommensurate transitions of quantum Hall stripe states
in double-quantum-well systems }
\author{R. C\^{o}t\'{e}}
\affiliation{D\'{e}partement de physique and CERPEMA, Universit\'{e} de Sherbrooke,
Sherbrooke, Qu\'{e}bec, Canada, J1K 2R1}
\author{H. A. Fertig}
\affiliation{Department of Physics and Astronomy, University of Kentucky, Lexington KY
40506-0055}
\author{J. Bourassa}
\affiliation{D\'{e}partement de physique and CERPEMA, Universit\'{e} de Sherbrooke,
Sherbrooke, Qu\'{e}bec, Canada, J1K 2R1}
\author{D. Bouchiha}
\affiliation{D\'{e}partement de physique and CERPEMA, Universit\'{e} de Sherbrooke,
Sherbrooke, Qu\'{e}bec, Canada, J1K 2R1}

\begin{abstract}
In higher Landau levels ($N>0$) and around filling factors $\nu =4N+1$, a
two-dimensional electron gas in a double-quantum-well system supports a
stripe groundstate in which the electron density in each well is spatially
modulated. When a parallel magnetic field is added in the plane of the
wells, tunneling between the wells acts as a spatially rotating effective
Zeeman field coupled to the ``pseudospins'' describing the well index of the
electron states. For small parallel fields, these pseudospins follow this
rotation, but at larger fields they do not, and a
commensurate-incommensurate transition results. Working in the Hartree-Fock
approximation, we show that the combination of stripes and commensuration in
this system leads to a very rich phase diagram. The parallel magnetic field
is responsible for oscillations in the tunneling matrix element that induce
a complex sequence of transitions between commensurate and incommensurate
liquid or stripe states. The homogeneous and stripe states we find can be
distinguished by their collective excitations and tunneling $I-V$, which we
compute within the time-dependent Hartree-Fock approximation.
\end{abstract}

\date{\today}
\maketitle

\section{Introduction}

In strong magnetic fields, the two-dimensional electron gas (2DEG) in a
double quantum well system (DQWS) is well known to support spontaneous
interlayer coherence if the interwell separation is smaller than some
critical value $d_{c}$. At filling factor $\nu =1$, the interlayer coherence
is reponsible for many interesting properties of the 2DEG such as a quantum
Hall state, a Goldstone mode \cite{fertiggoldstone} (in the absence of
tunneling) associated with the broken U(1) symmetry of the pseudospin order
in the ground state, which has been recently detected experimentally \cite%
{spielman}, and exotic topological quasiparticle excitations such as merons
and bimerons. When the magnetic field is inclined towards the plane of the
2DEG in the $\nu =1$ coherent state, a radical change in the behavior of the
transport gap was observed \cite{murphy} and explained as a
commensurate-incommensurate transition in the pseudospin order\cite%
{macdocomincom}. A review of these properties can be found in Ref.~%
\onlinecite{reviewcoherent}.

At bigger filling factors, in Landau levels $N>0$, it has been shown \cite%
{breydoublestripes} that the 2DEG in a DQWS can support states where both
the electronic densities and the interlayer coherence are spatially
modulated. Ref.~\onlinecite{breydoublestripes} discussed the ground state of
the DQWS at filling factors $\nu =4N+1$ when both wells have exactly the
same number of electrons using the Hartree-Fock approximation. They showed
that, for some range of interwell separations, a ``staggered'' stripe state
is lowest in energy. This state involves a unidirectional charge density
wave ordering of electrons in each well, such that the regions of high
electron density in one well are nearest to regions of low electron density
in the other. Stripe states in single well systems were first discussed by
Koulakov, Fogler, and Shklovskii and Moessner and Chalker \cite{stripesdebut}
and are believed to explain a large anisotropy observed in transport
experiments for quantum Hall systems in moderate magnetic fields, for which
the highest energy occupied Landau level is approximately half-filled \cite%
{expstripes}.

In a DQWS, stripe states are more complex and interesting than in the single
layer case because of the possibility that interlayer coherence may coexist
with charge ordering. In the staggered stripe state, interlayer coherence is
maintained only along linear regions at the edges of the stripes. These
``linear coherent regions'' (LCR's), whose width decreases with increasing
interlayer separation, support topological quasiparticle excitations closely
akin to sine-Gordon solitons \cite{breydoublestripes}. In the absence of
tunneling, C\^{o}t\'{e} and Fertig\cite{cotedoublestripes1} showed that this
staggered stripe state supports two Goldstone modes associated respectively
with broken translational symmetry and broken U(1) symmetry associated with
interlayer coherence.

In this paper, we address the questions of how the staggered stripe state
and its collective excitations are modified by the presence of a magnetic
field component parallel to the plane of the layers and how these changes
are reflected in the tunneling current. We recall \cite{reviewcoherent}
that, in the presence of a parallel magnetic field, the tunneling
hamiltonian can be written as an effective Zeeman coupling of the form%
\begin{equation}
H_{T}=-\int d\mathbf{r}\text{ }\mathbf{h}\left( \mathbf{r}\right) \cdot 
\mathbf{S}\left( \mathbf{r}\right) ,
\end{equation}%
where $\mathbf{S}\left( \mathbf{r}\right) \mathbf{\ }$is the pseudospin
density and $\mathbf{h}\left( \mathbf{r}\right) $ is a pseudo-magnetic field
that rotates in the $xy$ plane. In this paper, we choose $\mathbf{B}_{\Vert
} $ along the $y$ axis and use a gauge in which this pseudo-field is given
by 
\begin{equation}
\mathbf{h}\left( \mathbf{r}\right) =2t_{N}\left( Q\right) \left[ \cos \left(
Qx\right) \widehat{\mathbf{x}}-\sin \left( Qx\right) \widehat{\mathbf{y}}%
\right] ,
\end{equation}%
where $Q=d/\ell _{\Vert }^{2}$, with $d$ the separation between the wells,
and $\ell _{\Vert }=\sqrt{\hbar c/eB_{\Vert }}$ the magnetic length due to
the in-plane component $B_{\Vert }$ of the magnetic field. At filling factor 
$\nu =1$, the ground state at small layer separation and $B_{\Vert }=0$ is
well approximated by a state in which all pseudospins are oriented along the 
$x$ axis (we assume that the bare tunneling energy $t_{0}$ is not zero).
When the parallel magnetic field is turned on, the pseudospins rotate in the 
$xy$ plane, aligning with the local orientation of $\mathbf{h}\left( \mathbf{%
r}\right) $. This defines a uniform commensurate state, which we will call
the commensurate liquid (CL). When $B_{\Vert }$ increases, the rotation
period of the pseudospins decreases. This rotation is opposed by the Coulomb
exchange interaction that tends to keep peudospins parallel. Above some
critical $B_{\Vert }$, there is a phase transition to an incommensurate
state where the pseudospins no longer rotate with the in-plane field but
instead tend to align along some common direction as $B_{\Vert }$ increases
above the critical value. We call this state the incommensurate liquid (IL)
state. It is this change in the ground state of the 2DEG with $B_{\Vert }$
that is believed to be responsible for the radical change in the behavior of
the transport gap observed by Murphy et al. \cite{murphy}

In the experiment of Murphy \textit{et al.}, the system consisted of two
homogeneous two-dimensional electron gas (2DEG) with filling factor (in each
well) $\widetilde{\nu }=1/2$ and in Landau level $N=0$. The same
commensurate-incommensurate transition should occur in higher Landau levels
at total filling factors $\nu =4N+1$. In this case, however, we must
recognize that, for $d>d_{c}\left( N\right) $, the system is unstable to the
formation of stripes as was shown in Ref.~\onlinecite{cotedoublestripes1}.
In the pseudospin language, this staggered stripe state consists of
pseudospin modulation in the $xz$ plane. Applying a parallel magnetic field
will tend to induce a commensurate rotation of the pseudospins in the $xy$
plane in addition to the $xz$ modulations. In this paper, we call this phase
a commensurate stripe phase (CSP). If the parallel field is too strong, the
exchange energy cost of this commensurate rotation will be prohibitive and
the ground state will revert to a state with $xz$ modulations only as if
there were no tunneling and no parallel field. We call this other state an
incommensurate stripe phase (ISP). As $d$ and $B_{\Vert }$ are changed for a
given tunneling parameter, we thus expect a rich phase diagram with
transitions between the four possible states just introduced: CL, IL, CSP\
and ISP. (We cannot, of course, rule out the possibility of other ground
states with lower energy).

In higher Landau levels, the effect of $B_{\Vert }$ is more complex than for 
$N=0$. This is because the tunneling amplitude in the $N$th Landau level has
a non trivial $B_{\Vert }$ dependence, $t_{N}\left( Q\right)
=t_{0}e^{-Q^{2}\ell _{\bot }^{2}/4}L_{N}^{0}\left( \frac{Q^{2}\ell _{\bot
}^{2}}{2}\right) $, with $t_{0}$ the bare tunneling amplitude. For $N>0$,
this tunneling amplitude oscillates with wavevector $Q=d/\ell _{\Vert }^{2}$%
, becoming negative in some range of $B_{\Vert }$. One effect of these
oscillations is that $B_{\Vert }$ can induce not only one but several
(reentrant) commensurate-incommensurate transitions in the liquid or in the
stripe phases depending on the value of $N$. Figure 1 illustrates some
typical phase diagrams we find for several values of the bare tunneling
parameter $t_{0}$. These figures will be explained in more detail below.

\begin{figure}[tbp]
\includegraphics[width=8.25cm]{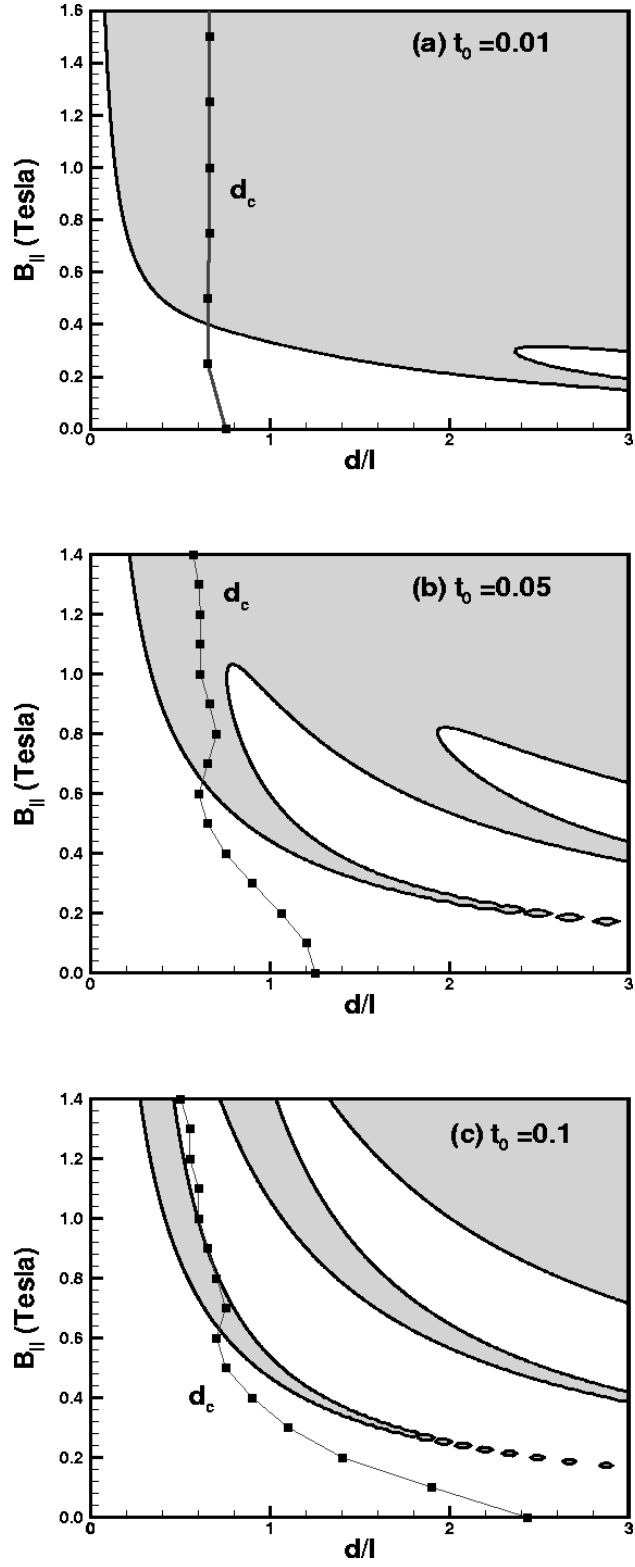}
\caption{Phase diagram of the 2DEG in Landau level $N=2$ in the $B_{\Vert
}-d/\ell _{\bot }$ plane for values of the bare tunneling parameters equal
to (a) $t_{0}/\left( e^{2}/\protect\kappa \ell \right) =0.01$, (b) $%
t_{0}/\left( e^{2}/\protect\kappa \ell \right) =0.05$ and (c) $t_{0}/\left(
e^{2}/\protect\kappa \ell \right) =0.1$. The areas in white(gray) represent
the commensurate(incommensurate) liquid state. The heavy line marked $d_{c}$
indicates the critical spacing above which one of the stripe state (either
commensurate or incommensurate) is lower in energy than both the CL and IL
phases.}
\label{fig1}
\end{figure}

Another effect of the parallel field is that, in the commensurate stripe
phase (CSP), the long-wavelength dispersion relations of the low-energy
phonon and pseudospin wave modes will be strongly dependent on $B_{\Vert }$
through $t_{N}\left( Q\right) $. We find that the gap in the pseudospin wave
mode (which is related to the broken U(1) symmetry of the pseudospin order)
and in the phonon mode (at finite wavevector) follows the oscillations of
the tunneling amplitude $t_{N}\left( Q\right) $. In particular, the phonon
and pseudospin wave modes become essentially gapless whenever the parallel
field is such that $t_{N}\left( Q\right) $ is close to one of its $N$
zeroes. Such behavior is not seen in the incommensurate stripe phase (ISP)
since the properties of this state are independent of $t_{N}\left( Q\right) $
and so of $B_{\Vert }$. We thus find that the CSP and ISP are clearly
distinguishable by their low-energy collective modes. Figure 2 illustrates
these modes as computed using the time-dependent Hartree-Fock approximation,
as explained in more detail below.

\begin{figure}[tbp]
\includegraphics[width=8.25cm]{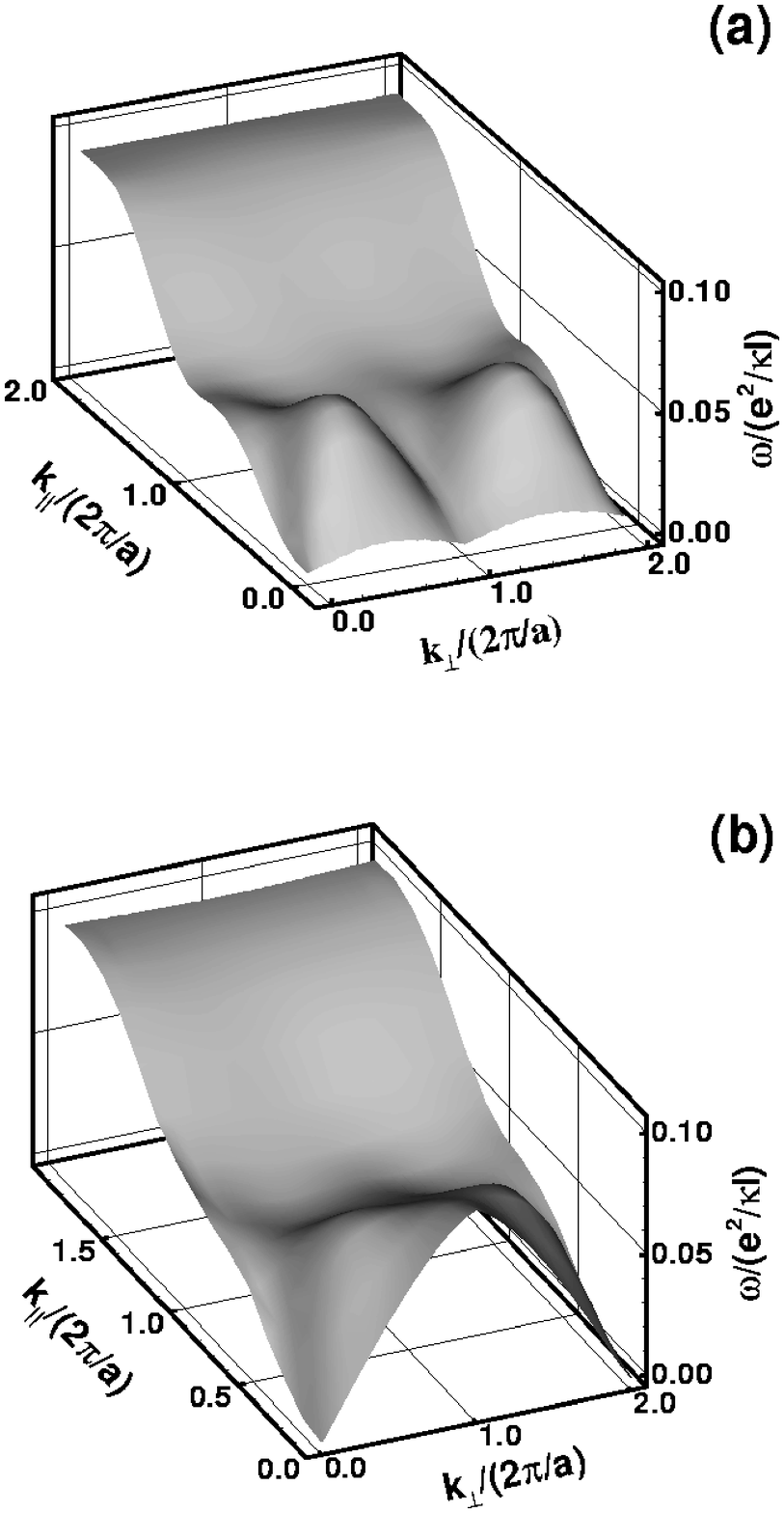}
\caption{Dispersion relations of the low-energy modes in (a) the ISP and (b)
the CSP for $N=2,t_{0}/\left( e^{2}/\protect\kappa \ell \right) =0.01$ and $%
B_{\Vert }=0$ T.}
\label{fig2}
\end{figure}

One way to probe these states is via interlayer tunneling. For very small
values of $t_{0}$, the tunneling $I-V$ has been argued \cite{tunneltheory}
and shown \cite{spielman} to probe the Goldstone mode dispersion of the
(spontaneously) coherent state as the parallel field is adjusted. An
analogous effect occurs in the stripe state of this system, as illustrated
in Fig.~3. A clear signature of the existence of stripes can be found by
noting the non-monotonic behavior of the tunneling peak as a function of $%
B_{\Vert }$. In particular, one finds this peak at zero bias voltage for $%
B_{\Vert }=nhc/ade$ where $a$ is the separation between the stripes in a
given well and $n=0,1,2,\ldots $. This is a reflection of the periodicity of
the collective mode spectrum illustrated in Fig.~2(a), and directly
demonstrates the existence of stripes in the groundstate.

\begin{figure}[tbp]
\includegraphics[width=8.5cm]{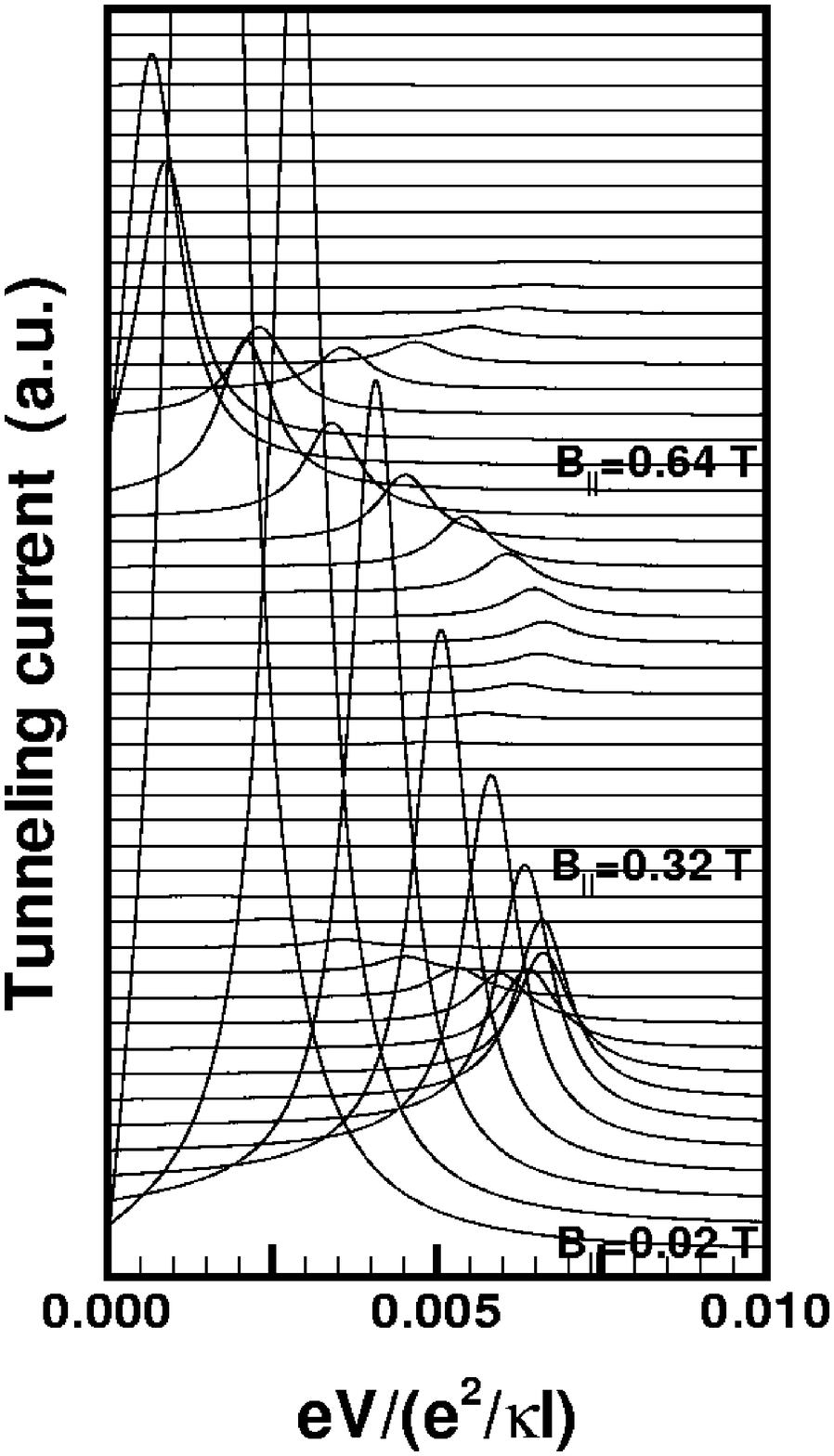}
\caption{Variation of the tunneling current $I\left( t\right) $ with the
potential biais $eV$ for several values of the parallel magnetic field in
the CSP. The tunneling current is calculated by using the first term in Eq.~(%
\ref{tunnel1}) and with $N=2,t_{0}/\left( e^{2}/\protect\kappa \ell \right)
=0.01$ and $d/\ell _{\bot }=1.4$. Note that $B_{\Vert }=0.325$ T corresponds
approximately to $Q\ell _{\bot }=2\protect\pi /a$ where $a$ is the
separation between the stripes in a given well.}
\label{fig3}
\end{figure}

This paper is organized as follow. In Sec. II, we derive the tunneling
Hamiltonian in the presence of the parallel magnetic field and explain the
origin of the oscillations in $t_{N}\left( Q\right) $. In Sec. III, we
derive the Hartree-Fock equations of motion for the single-particle Green's
function from which we compute the density and pseudospin density as well as
the energy of the commensurate and incommensurate stripe states. The energy
of these phases are compared with those of the commensurate and
incommensurate liquid phases in Sec. IV to obtain the phase diagram, in
Landau level $N=2$, for various parameters $B_{\Vert },t_{0},$ and $d$. We
briefly review, in Sec. V, the formalism necessary to compute the
susceptibilities and dispersion relations in the time-dependent Hartree-Fock
approximation (TDHFA) and present our numerical results in various region of
the phase diagram studied in Sec. IV. Section VI shows how the tunneling
current, in the stripe states, is affected by the parallel magnetic field.
We conclude in Sec. VII.

\section{Tunneling Hamiltonian}

We consider an unbiased symmetric DQWS in a tilted magnetic field $\mathbf{B}%
=B_{\Vert }\widehat{\mathbf{y}}+B_{\bot }\widehat{\mathbf{z}}$ at total
filling factor $\nu =4N+1$ where $N=0,1,2,...$ is the Landau level index. To
simplify the analysis, we make several approximations. We retain only one
electric subband of the DQWS, and assume that $g^{\ast }\mu _{B}B>>\Delta
_{SAS}$ where $\Delta _{SAS}$ is the symmetric to antisymmetric gap so that
the real spins are completely frozen. Furthermore, we consider the lower
Landau levels as filled and inert and neglect any Landau level mixing.
Considering very narrow wells and treating the tunneling term in a
tight-binding approximation, the non-interacting Hamiltonian of the 2DEG can
be written, in the gauge where $\mathbf{A}=\left( 0,B_{\bot }x,-B_{\Vert
}x\right) $, as%
\begin{equation}
H_{0}=\sum_{X,j}E_{0}c_{X,j}^{\dagger }c_{X,j}-\sum_{X}t_{N}\left( Q\right)
\left( e^{iQX}c_{X,R}^{\dagger }c_{X,L}+e^{-iQX}c_{X,L}^{\dagger
}c_{X,R}\right) ,  \label{tunnel3}
\end{equation}%
where $E_{0}$ is the electron energy in each isolated well and $c_{\alpha
,X}^{\dagger }$ is the creation operator for an electron in Landau level $N$%
, guiding center index $X$, and well index $j=R,L$. The electronic
wavefunctions are given by 
\begin{equation}
\psi _{N,X,j}(\mathbf{r})=\frac{1}{\sqrt{L_{y}}}e^{-iXy/\ell ^{2}}\varphi
_{N}\left( x-X\right) \chi _{j}(z),
\end{equation}%
where $\chi _{j}(z)$ is the envelope wave function of the lowest-energy
electric subband centered on the right or left well and $\varphi _{N}\left(
x\right) $ is an eigenfunction of the one-dimensional harmonic oscillator.
In the tight-binding approximation, the tunneling amplitude, $t_{N}\left(
Q\right) $, is given by%
\begin{equation}
t_{N}\left( Q\right) =-t_{0}\int_{-\infty }^{+\infty }dx\varphi _{N}\left(
x-X\right) e^{iQx}\varphi _{N}\left( x-X\right) =t_{0}e^{-Q^{2}\ell _{\bot
}^{2}/4}L_{N}^{0}\left( \frac{Q^{2}\ell _{\bot }^{2}}{2}\right) ,
\end{equation}%
where $t_{0}$ is the bare tunneling amplitude (defined as a positive
quantity), $L_{N}^{0}\left( x\right) $ is a generalized Laguerre's
polynomial and the wavevector and $Q=d/\ell _{\Vert }^{2}$ where $\ell
_{\Vert }=\sqrt{\hslash c/eB_{_{\Vert }}}$ is the magnetic length associated
with the parallel magnetic field.

If $N>0$, the tunneling amplitude oscillates with the strength of the
parallel field and/or the separation between the wells \cite{macdotunnel}.
(In this paper, we neglect the further dependency of $t_{0}$ on $d$ due to
the change in the overlap of the electric subband wavefunctions $\chi
_{j}(z) $ of the two wells). We remark that the first zero of $t_{N}\left(
Q\right) $ appears at relatively small parallel field. For example, since $%
2\pi n_{0}\ell _{\bot }^{2}=\nu $ where $n_{0}$ is the total areal density
of the electrons in the DQWS, the wavevector $Q\ell _{\bot }=\nu \left(
d/\ell _{\bot }\right) \left( e/n_{0}hc\right) B_{\Vert },$ and so $d/\ell
_{\bot }=d\sqrt{2\pi n_{0}/\nu }$. If we assume a typical total electronic
density of $n_{0}=1\times 10^{11}$ cm$^{-2}$, we have, with $B_{\Vert }$ in
Tesla, $Q\ell _{\bot }=0.242\left( \nu \right) \left( d/\ell _{\bot }\right)
B_{\Vert }$. On the other hand, at this density the separation between the
wells can be expressed as $\left( d/\ell _{\bot }\right) =0.793d(\mathrm{100}
$\AA $)/\sqrt{\nu }$. For total filling factor $\nu =1$, we get $d/\ell
_{\bot }=1.6$ for a typical interwell separation of $200$ \AA . This value
decreases as $N$ increases at fixed total density. Figure 4 shows the
behavior of $t_{N}\left( Q\right) $ for $d/\ell _{\bot }=1.0$, $%
n_{0}=1\times 10^{11}$ cm$^{-2}$ and filling fractions $\nu =1,5,9$
corresponding to $N=0,1,2$. We see from this figure that for $N=2$, $%
t_{N}(Q) $ vanishes near $B_{\Vert }\approx 0.5$ T and $1.2$ T, and that a
parallel field of $\approx 3$ T is required to reach the asymptotic behavior
of the tunneling amplitude.

\begin{figure}[tbp]
\includegraphics[width=8.5cm]{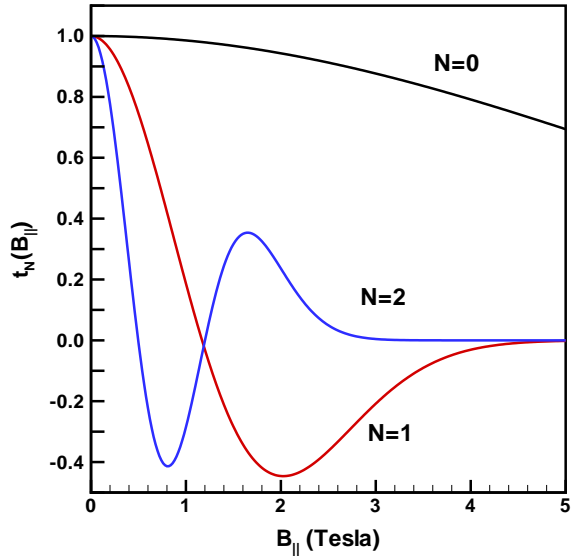}
\caption{Variation of the tunneling amplitude $t_{N}\left( Q\right) $ with
the parallel magnetic field for filling fractions $\protect\nu =1,5,9$
corresponding to Landau levels $N=0,1,2$ and for $d/\ell _{\bot }=1.0$. }
\label{fig4}
\end{figure}

\section{Hartree-Fock Equations of Motion\qquad}

We describe the various ordered coherent states of the electrons in the
partially filled Landau level $N$ by the set of average values $\left\{
\left\langle \rho _{i,j}(\mathbf{q})\right\rangle \right\} $ where $\rho
_{i,j}(\mathbf{q})$ is an operator that we define\cite{cotemethode} by 
\begin{equation}
\rho _{i,j}(\mathbf{q})=\frac{1}{N_{\varphi }}%
\sum_{X}e^{-iq_{x}X+iq_{x}q_{y}\ell ^{2}/2}\ c_{i,X}^{\dagger
}c_{j,X-q_{y}\ell ^{2}},
\end{equation}%
where $N_{\varphi }$ is the Landau level degeneracy. At a semiclassical
level, the $\left\langle \rho _{i,i}(\mathbf{q})\right\rangle ^{\prime }s$
can be thought of as the Fourier transform of the density of guiding centers
of the cyclotron orbits while the off diagonal terms $\left\langle \rho
_{i\neq j}(\mathbf{q})\right\rangle $ describe quantum coherence between the
two 2DEG. The average values $\left\langle \rho _{i,j}(\mathbf{q}%
)\right\rangle $ are obtained by computing the single-particle Green's
function 
\begin{equation}
G_{i,j}\left( X,X^{\prime },\tau \right) =-\left\langle Tc_{i,X}\left( \tau
\right) c_{j,X^{\prime }}^{\dagger }\left( 0\right) \right\rangle ,
\end{equation}%
whose Fourier transform we define as 
\begin{equation}
G_{i,j}\left( \mathbf{q,}\tau \right) =\frac{1}{N_{\phi }}\sum_{X,X^{\prime
}}e^{-\frac{i}{2}q_{x}\left( X+X^{\prime }\right) }\delta _{X,X^{\prime
}-q_{y}l^{2}}G_{i,j}\left( X,X^{\prime },\tau \right) ,
\end{equation}%
so that $G_{i,j}\left( \mathbf{q,}\tau =0^{-}\right) =\left\langle \rho
_{j,i}\left( \mathbf{q}\right) \right\rangle $.

Using the tunneling Hamiltonian $H_{0}$ [Eq.~(\ref{tunnel3})] and working in
the Hartree-Fock approximation, we find that the single-particle Green's
functions for the DQWS obey the system of equations

\begin{gather}
\left[ i\hslash \omega _{n}-\left( E_{0}-\mu \right) \right] G_{R,R}\left( 
\mathbf{q},\omega _{n}\right) +t^{\ast }\left( \mathbf{q}\right)
G_{L,R}\left( \mathbf{q-Q},\omega _{n}\right)  \label{system1debut1} \\
-\sum_{\mathbf{q}^{\prime }}U_{R,R}\left( \mathbf{q,q}^{\prime }\right)
G_{R,R}\left( \mathbf{q}^{\prime },\omega _{n}\right) -\sum_{\mathbf{q}%
^{\prime }}U_{R,L}\left( \mathbf{q,q}^{\prime }\right) G_{L,R}\left( \mathbf{%
q}^{\prime },\omega _{n}\right) =\hslash \delta _{\mathbf{q},0},  \notag
\end{gather}

\begin{gather}
\left[ i\hslash \omega _{n}-\left( E_{0}-\mu \right) \right] G_{R,L}\left( 
\mathbf{q},\omega _{n}\right) +t^{\ast }\left( \mathbf{q}\right)
G_{L,L}\left( \mathbf{q-Q},\omega _{n}\right) \\
-\sum_{\mathbf{q}^{\prime }}U_{R,R}\left( \mathbf{q,q}^{\prime }\right)
G_{R,L}\left( \mathbf{q}^{\prime },\omega _{n}\right) -\sum_{\mathbf{q}%
^{\prime }}U_{R,L}\left( \mathbf{q,q}^{\prime }\right) G_{L,L}\left( \mathbf{%
q}^{\prime },\omega _{n}\right) =0,  \notag
\end{gather}%
\begin{gather}
\left[ i\hslash \omega _{n}-\left( E_{0}-\mu \right) \right] G_{L,R}\left( 
\mathbf{q},\omega _{n}\right) +t\left( \mathbf{q}\right) G_{R,R}\left( 
\mathbf{q+Q},\omega _{n}\right) \\
-\sum_{\mathbf{q}^{\prime }}U_{L,R}\left( \mathbf{q,q}^{\prime }\right)
G_{R,R}\left( \mathbf{q}^{\prime },\omega _{n}\right) -\sum_{\mathbf{q}%
^{\prime }}U_{L,L}\left( \mathbf{q,q}^{\prime }\right) G_{L,R}\left( \mathbf{%
q}^{\prime },\omega _{n}\right) =0,  \notag
\end{gather}%
\begin{gather}
\left[ i\hslash \omega _{n}-\left( E_{0}-\mu \right) \right] G_{L,L}\left( 
\mathbf{q},\omega _{n}\right) +t\left( \mathbf{q}\right) G_{R,L}\left( 
\mathbf{q+Q},\omega _{n}\right)  \label{system1fin} \\
-\sum_{\mathbf{q}^{\prime }}U_{L,R}\left( \mathbf{q,q}^{\prime }\right)
G_{R,L}\left( \mathbf{q}^{\prime },\omega _{n}\right) -\sum_{\mathbf{q}%
^{\prime }}U_{L,L}\left( \mathbf{q,q}^{\prime }\right) G_{L,L}\left( \mathbf{%
q}^{\prime },\omega _{n}\right) =\hslash \delta _{\mathbf{q},0}.  \notag
\end{gather}%
The tunneling term $t\left( \mathbf{q}\right) $ is, in our particular
geometry where the parallel magnetic field is along the $y$ axis, given by 
\begin{equation}
t\left( \mathbf{q}\right) =t_{N}\left( Q\right) e^{iQq_{y}l_{\bot }^{2}/2}.
\end{equation}%
When $Q=0$, Eqs. (\ref{system1debut1})-(\ref{system1fin}) is the system of
equations we studied in Ref.~\onlinecite{cotedoublestripes1}. The Coulomb
interactions and averaged densities in this system of equations produce the
mean field potentials 
\begin{equation}
U_{R,R}\left( \mathbf{q,q}^{\prime }\right) =\left( \frac{e^{2}}{\kappa
l_{\bot }}\right) \left[ H\left( \mathbf{q}-\mathbf{q}^{\prime }\right)
-X\left( \mathbf{q}-\mathbf{q}^{\prime }\right) \right] \left\langle \rho
_{R,R}\left( \mathbf{q-q}^{\prime }\right) \right\rangle \gamma _{\mathbf{q},%
\mathbf{q}^{\prime }}+\left( \frac{e^{2}}{\kappa l_{\bot }}\right) 
\widetilde{H}\left( \mathbf{q}-\mathbf{q}^{\prime }\right) \left\langle \rho
_{L,L}\left( \mathbf{q-q}^{\prime }\right) \right\rangle \gamma _{\mathbf{q},%
\mathbf{q}^{\prime }},  \notag
\end{equation}%
\begin{equation}
U_{L,L}\left( \mathbf{q,q}^{\prime }\right) =\left( \frac{e^{2}}{\kappa
l_{\bot }}\right) \left[ H\left( \mathbf{q}-\mathbf{q}^{\prime }\right)
-X\left( \mathbf{q}-\mathbf{q}^{\prime }\right) \right] \left\langle \rho
_{L,L}\left( \mathbf{q-q}^{\prime }\right) \right\rangle \gamma _{\mathbf{q},%
\mathbf{q}^{\prime }}+\left( \frac{e^{2}}{\kappa l_{\bot }}\right) 
\widetilde{H}\left( \mathbf{q}-\mathbf{q}^{\prime }\right) \left\langle \rho
_{R,R}\left( \mathbf{q-q}^{\prime }\right) \right\rangle \gamma _{\mathbf{q},%
\mathbf{q}^{\prime }},  \notag
\end{equation}%
\begin{equation}
U_{R,L}\left( \mathbf{q,q}^{\prime }\right) =-\left( \frac{e^{2}}{\kappa
l_{\bot }}\right) \widetilde{X}\left( \mathbf{q}-\mathbf{q}^{\prime }\right)
\left\langle \rho _{L,R}\left( \mathbf{q-q}^{\prime }\right) \right\rangle
\gamma _{\mathbf{q},\mathbf{q}^{\prime }},
\end{equation}%
\begin{equation}
U_{L,R}\left( \mathbf{q,q}^{\prime }\right) =-\left( \frac{e^{2}}{\kappa
l_{\bot }}\right) \widetilde{X}\left( \mathbf{q}-\mathbf{q}^{\prime }\right)
\left\langle \rho _{R,L}\left( \mathbf{q-q}^{\prime }\right) \right\rangle
\gamma _{\mathbf{q},\mathbf{q}^{\prime }},
\end{equation}%
where 
\begin{equation}
\gamma _{\mathbf{q},\mathbf{q}^{\prime }}=e^{-i\left( \mathbf{q}\times 
\mathbf{q}^{\prime }\right) \cdot \widehat{\mathbf{z}}l_{\bot }^{2}/2}.
\end{equation}%
The definitions of the coulomb interactions $H,X,\widetilde{H}$ and $%
\widetilde{X}$ were given in Ref.~\onlinecite{cotedoublestripes1}. They are
respectively the Fourier transforms of the intrawell Hartree and Fock
interactions and the interwell Hartree and Fock interactions.

Once the densities are extracted from the single-particle Green's functions,
the Hartree-Fock energy per electron can be written as%
\begin{eqnarray}
\frac{E_{HF}}{N_e} &=&-\frac{2}{\nu }t_{N}\left( Q\right) \func{Re}\left[
\left\langle \rho _{R,L}\left( \mathbf{Q}\right) \right\rangle \right]
\label{energystripesincom} \\
&&+\frac{1}{2\nu }\left( \frac{e^{2}}{\kappa l_{\bot }}\right) \sum_{\mathbf{%
q}\neq 0}H\left( \mathbf{q}\right) \left[ \left| \left\langle \rho
_{R,R}\left( \mathbf{q}\right) \right\rangle \right| ^{2}+\left|
\left\langle \rho _{L,L}\left( \mathbf{q}\right) \right\rangle \right| ^{2}%
\right]  \notag \\
&&-\frac{1}{2\nu }\left( \frac{e^{2}}{\kappa l_{\bot }}\right) \sum_{\mathbf{%
q}}X\left( \mathbf{q}\right) \left[ \left| \left\langle \rho _{R,R}\left( 
\mathbf{q}\right) \right\rangle \right| ^{2}+\left| \left\langle \rho
_{L,L}\left( \mathbf{q}\right) \right\rangle \right| ^{2}\right]  \notag \\
&&+\frac{1}{\nu }\left( \frac{e^{2}}{\kappa l_{\bot }}\right) \sum_{\mathbf{q%
}\neq 0}\widetilde{H}\left( \mathbf{q}\right) \left[ \left\langle \rho
_{R,R}\left( -\mathbf{q}\right) \right\rangle \left\langle \rho _{L,L}\left( 
\mathbf{q}\right) \right\rangle \right]  \notag \\
&&-\frac{1}{\nu }\left( \frac{e^{2}}{\kappa l_{\bot }}\right) \sum_{\mathbf{q%
}}\widetilde{X}\left( \mathbf{q}\right) \left| \left\langle \rho
_{L,R}\left( \mathbf{q}\right) \right\rangle \right| ^{2}.  \notag
\end{eqnarray}

The $\left\langle \rho _{i,j}\left( \mathbf{q}\right) \right\rangle ^{\prime
}s$ are found by solving Eqs. (10)-(13), and using Eqs. (7)-(9) to extract
them from the resulting Green's functions. Since the potentials entering the
former set of equations involve the $\left\langle \rho _{i,j}\left( \mathbf{q%
}\right) \right\rangle ^{\prime }s$, this procedure must be iterated until
convergence is achieved. (The numerical scheme is described in more detail
in Ref.~\onlinecite{cotemethode}.) We next describe the results of such
calculations for several possible states, and discuss the phase diagram of
the system.

\section{Non-homogeneous Hartree-Fock States in Higher Landau Levels}

In Refs.~\onlinecite{breydoublestripes} and \onlinecite{cotedoublestripes1},
it was found, for $B_{\Vert }=0$, that, with the exception of the $N=0$
Landau level, the 2DEG ground state first evolves from a uniform coherent
state (UCS)\ to a unidirectional CDW state, and then undergoes a first order
transition to a modulated stripe phase as $d/\ell _{\bot }$ is increased
from zero. Wigner crystal (WC) states were found to be higher in energy than
the unidirectional modulated states for $N>0$. In the $N=0$ Landau level,
the evolution was shown to be from a uniform coherent state to a CDW state
and then to a Wigner crystal state at large interwell separation. The
coherent states (either CDW or WC) are always lower in energy than their
incoherent counterparts.

In this paper, we consider the consequences to the stability of these
unidirectionally ordered states of adding a parallel magnetic field. In view
of our earlier results, we restrict ourselves to higher Landau levels $N>0$
and consider only coherent states. (Incoherent states are insensitive to the
parallel field). Our filling factor is $\nu =4N+1$ and we consider four
possible groundstates. These states are best described in a pseudospin
language where the right (left) well electronic state is associated with a
pseudospin up (down) state. The $\rho _{i,j}$ operators defined above can be
mapped into the density and pseudospin density operators by the relations 
\begin{eqnarray}
n\left( \mathbf{q}\right) &=&\frac{1}{2}\left[ \rho _{R,R}\left( \mathbf{q}%
\right) +\rho _{L,L}\left( \mathbf{q}\right) \right] , \\
\rho _{x}\left( \mathbf{q}\right) &=&\frac{1}{2}\left[ \rho _{R,L}\left( 
\mathbf{q}\right) +\rho _{L,R}\left( \mathbf{q}\right) \right] , \\
\rho _{y}\left( \mathbf{q}\right) &=&\frac{1}{2i}\left[ \rho _{R,L}\left( 
\mathbf{q}\right) -\rho _{L,R}\left( \mathbf{q}\right) \right] , \\
\rho _{z}\left( \mathbf{q}\right) &=&\frac{1}{2}\left[ \rho _{R,R}\left( 
\mathbf{q}\right) -\rho _{L,L}\left( \mathbf{q}\right) \right] .
\end{eqnarray}%
The tunneling hamiltonian of Eq.~(\ref{tunnel3}) can be written as an
effective ``Zeeman'' coupling%
\begin{equation}
H_{T}=-\sum_{X}\mathbf{h}\left( X\right) \cdot \mathbf{\rho }\left( X\right)
,  \label{tunnelpseudo}
\end{equation}%
where $\mathbf{\rho }\left( X\right) $ and $n\left( X\right) $ are the
(guiding center-) spin density and density operators respectively. The field 
$\mathbf{h}\left( X\right) $ plays the role of a pseudo-magnetic field that
lies in the plane of the 2DEG and rotates in the space of the $X$ coordinate
according to 
\begin{equation}
\mathbf{h}\left( X\right) =2t_{N}\left( Q\right) \left[ \cos \left(
QX\right) \widehat{\mathbf{x}}-\sin \left( QX\right) \widehat{\mathbf{y}}%
\right] ,  \label{pseudofield}
\end{equation}%
with a period $\lambda =2\pi /Q=2\pi \ell _{\Vert }^{2}/d$. (The minus sign
in Eq.~(\ref{pseudofield}) is due to our particular convention for
associating the $R$ and $L$ wells with the up and down spins).

When the magnetic field is tilted towards the plane of the 2DEG (at constant 
$B_{\bot }$ to keep the filling fraction constant), the rotation period $%
\lambda $ of the pseudospins decreases. In a non-interacting picture, the
pseudospins would align with the local field $\mathbf{h}\left( X\right) $ to
minimize the tunneling energy. When electron-electron interactions are taken
into account, however, the pseudospins interact through an effective
exchange interaction\cite{macdocomincom} that favors a parallel
configuration of the pseudospins. At small parallel magnetic field, the
pseudospins can follow the effective Zeeman field. The resulting state is
called the commensurate liquid (CL). Above some critical parallel magnetic
field, $B_{\Vert ,c}$, the exchange energy cost of the rotation is too high,
and an incommensurate liquid (IL) state where the pseudospins are spatially
uniform becomes lower in energy. We note that this incommensurate state is
never the true ground state: at a field slightly lower than $B_{\Vert ,c}$,
the 2DEG has a transition into a soliton lattice state (SLS) where kinks
(slips of $2\pi $ in the phase of the rotating pseudospins) form a lattice
in the direction perpendicular to the parallel magnetic field with a period $%
\lambda _{s}\left( Q\right) $. This SLS continuously evolves into the IL as
the parallel field is increased. We do not explicitly treat the SLS in our
work since it is significantly different than the IL only in a small range
of $Q$ (see, for example, Ref.~\onlinecite{cotesls}). We will thus refer to
the commensurate-incommensurate transition as to the transition between the
CL and the IL.

The CL and IL states are spatially homogeneous; their orderings are
characterized by the behavior of $\left\langle \rho _{i\neq j}(\mathbf{q}%
)\right\rangle $ but have no non-vanishing values of $\left\langle \rho
_{ii}(\mathbf{q}\neq 0)\right\rangle $. In high Landau levels, stripe states
may form \cite{breydoublestripes,stripesdebut} which have guiding center
densities oscillating spatially between the wells. It was found in Ref. %
\onlinecite{breydoublestripes} that for $d/\ell _{\bot }$ not too small, the
edges of the stripes in each well will admix to form linear coherent regions
(LCR's), and that in the pseudospin language such states are akin to helical
ferromagnets \cite{cotedoublestripes1}. In the presence of a parallel field,
the pseudospin ordering in the LCR's may or may not follow the effective
Zeeman field, leading to a commensurate stripe phase (CSP) and an
incommensurate stripe phase (ISP) .

\subsection{Commensurate Liquid (CL)}

The commensurate liquid at $\nu =1$ and with $t_{N}\left( Q\right) >0$ is
defined by homogeneous and equal densities of electrons in both wells and by
a rotation of the pseudospins in the $xy$ plane with a single period $%
\lambda =2\pi \ell _{\Vert }^{2}/d$. The order parameters of this state are
given by 
\begin{eqnarray}
\left\langle \rho _{R,R}\left( 0\right) \right\rangle &=&\left\langle \rho
_{L,L}\left( 0\right) \right\rangle =\frac{1}{2}, \\
\left\langle \rho _{R,L}\left( -\mathbf{Q}\right) \right\rangle
&=&\left\langle \rho _{L,R}\left( \mathbf{Q}\right) \right\rangle =\frac{1}{2%
}.
\end{eqnarray}%
All other $\rho $'s vanish in this state. The energy per particle $%
E_{CL}=E_{HF}/N_e$ ($N_e$ is the number of electrons in the partially filled
level) is simply 
\begin{equation}
E_{CL}=-\left| t_{N}\left( Q\right) \right| -\frac{1}{4}\left( \frac{e^{2}}{%
\kappa l_{\bot }}\right) \left[ X\left( 0\right) +\widetilde{X}\left( 
\mathbf{Q}\right) \right] .  \label{energycl}
\end{equation}%
The absolute value in Eq.~(\ref{energycl}) is necessary since, when $%
t_{N}\left( Q\right) <0$, the pseudo-magnetic field reverses direction and
we must have $\left\langle \rho _{R,L}\left( -\mathbf{Q}\right)
\right\rangle =-\frac{1}{2}.$

\subsection{Incommensurate Liquid (IC)}

The incommensurate liquid is a state with equal and homogeneous densities in
both wells and with the pseudospins all aligned in the same direction which
we take here as the $x$ axis. The only non-vanishing order parameters are

\begin{eqnarray}
\left\langle \rho _{R,L}\left( 0\right) \right\rangle ,\left\langle \rho
_{L,R}\left( 0\right) \right\rangle &=&\frac{1}{2}, \\
\left\langle \rho _{R,R}\left( 0\right) \right\rangle ,\left\langle \rho
_{L,L}\left( 0\right) \right\rangle &=&\frac{1}{2},
\end{eqnarray}%
and the Hartree-Fock energy per particle is 
\begin{equation}
E_{IC}=-\frac{1}{4}\left( \frac{e^{2}}{\kappa l_{\bot }}\right) \left[
X\left( 0\right) +\widetilde{X}\left( 0\right) \right] ,  \label{energyil}
\end{equation}%
which does not depend on the tunneling term or the parallel magnetic field.

\subsection{Commensurate Stripe Phase (CSP)}

In what follows, the parallel magnetic field is assumed to be in the $%
\widehat{\mathbf{y}}$ direction so that $\mathbf{Q}=Q_{0}\widehat{\mathbf{x}}
$. The coherent stripes are defined by charge and spin modulations with
reciprocal lattice vectors $\mathbf{G}=\frac{2\pi }{a}n\widehat{\mathbf{e}}$
where $n=0,\pm 1,\pm 2,...$ and $\widehat{\mathbf{e}}$ is a unit vector in a
direction making an angle $\alpha $ with $\mathbf{Q}$ and the $x$ axis. In
the absence of the parallel magnetic field, the coherent stripes would be
described by the order parameters $\left\{ \left\langle \rho _{i,j}\left( 
\mathbf{G}\right) \right\rangle \right\} $. With the field, the situation is
slightly more complex since the tunneling term forces $\left\langle \rho
_{R,L}\left( \mathbf{Q}\right) \right\rangle \neq 0$ and in general $\mathbf{%
Q\notin }\left\{ \mathbf{G}\right\} $.

To apply our equation of motion method, we need a single set of wavevectors
to describe the modulated state. This can be achieved by considering a
solution for the transverse part of the spins, of the form

\begin{equation}
\left\langle \rho _{R,L}\left( \mathbf{r}\right) \right\rangle =e^{-i\mathbf{%
Q}\cdot \mathbf{r}}\left\langle \widetilde{\rho }_{R,L}\left( \mathbf{r}%
\right) \right\rangle ,  \label{rogcsp}
\end{equation}%
where $\left\langle \widetilde{\rho }_{R,L}\left( \mathbf{r}\right)
\right\rangle $ has the periodicity of the stripe state [i.e., $\left\langle 
\widetilde{\rho }_{R,L}\left( \mathbf{r}\right) \right\rangle \neq 0$ for $%
\mathbf{{q}\in \left\{ G\right\} }$]. The sole effect of the rotating field
is to force a rotation of the pseudospins in the $xy$ plane in the stripe
state which itself has the pseudospins rotating in the $xz$ plane. We call
this solution the commensurate stripe phase (CSP). Finding the order
parameters is most easily accomplished by going into a spatially rotating
frame where the $xy$ pseudospins are aligned. We define the average values
of the operators in the rotating frame by 
\begin{eqnarray}
\left\langle \widetilde{\rho }_{R,L}\left( \mathbf{q}\right) \right\rangle
&\equiv &\left\langle \rho _{R,L}\left( \mathbf{q-Q}\right) \right\rangle ,
\\
\left\langle \widetilde{\rho }_{L,R}\left( \mathbf{q}\right) \right\rangle
&\equiv &\left\langle \rho _{L,R}\left( \mathbf{q+Q}\right) \right\rangle .
\end{eqnarray}%
Because the system of equations (\ref{system1debut1})-(\ref{system1fin})
contains the factor $\gamma _{\mathbf{q},\mathbf{q}^{\prime }}$ , making the
substitution $\mathbf{q}\rightarrow \mathbf{q}\pm \mathbf{Q}$ introduces
additional phase factors in the equations. We can preserve the structure of
this system of equations in the rotating frame if we also redefine the
diagonal order parameters as 
\begin{eqnarray}
\left\langle \widetilde{\rho }_{m}^{R,R}\left( \mathbf{q}\right)
\right\rangle &\equiv &\gamma ^{\ast }\left( \mathbf{q}\right) \left\langle
\rho _{m}^{R,R}\left( \mathbf{q}\right) \right\rangle , \\
\left\langle \widetilde{\rho }_{m}^{L,L}\left( \mathbf{q}\right)
\right\rangle &\equiv &\gamma \left( \mathbf{q}\right) \left\langle \rho
_{m}^{L,L}\left( \mathbf{q}\right) \right\rangle ,
\end{eqnarray}%
where%
\begin{equation}
\gamma \left( \mathbf{q}\right) \equiv e^{i\mathbf{q}\times \mathbf{Q}%
l_{\bot }^{2}/2}=e^{-iq_{y}Ql_{\bot }^{2}/2}.
\end{equation}%
The CSP thus occurs when $\left\langle \widetilde{\rho }_{m}^{i,j}\left( 
\mathbf{q}\right) \right\rangle $ are non-vanishing for $\mathbf{{q}\in
\left\{ G\right\} }$.

To carry out the computation, it is convenient to define a $2\times 2$
Green's function matrix, with matrix elements 
\begin{eqnarray}
\widetilde{G}_{R,L}\left( \mathbf{q},\omega _{n}\right) &\equiv
&G_{R,L}\left( \mathbf{q+Q},\omega _{n}\right) , \\
\widetilde{G}_{L,R}\left( \mathbf{q},\omega _{n}\right) &\equiv
&G_{L,R}\left( \mathbf{q-Q},\omega _{n}\right) , \\
\widetilde{G}_{R,R}\left( \mathbf{q},\omega _{n}\right) &\equiv &\gamma
^{\ast }\left( \mathbf{q}\right) G_{R,R}\left( \mathbf{q},\omega _{n}\right)
, \\
\widetilde{G}_{L,L}\left( \mathbf{q},\omega _{n}\right) &\equiv &\gamma
\left( \mathbf{q}\right) G_{L,L}\left( \mathbf{q},\omega _{n}\right) .
\end{eqnarray}%
Using the fact that $\mathbf{Q}=Q\widehat{\mathbf{x}}$, we have 
\begin{equation}
t\left( \mathbf{q\pm Q}\right) =t\left( \mathbf{q}\right) .
\end{equation}%
With these definitions and identities, the system of equations for the
single-particle Green's function in the rotating frame is%
\begin{gather}
\left[ i\omega _{n}-\frac{1}{\hslash }\left( E_{0}-\mu \right) \right] 
\widetilde{G}_{R,R}\left( \mathbf{q},\omega _{n}\right) +\frac{1}{\hslash }%
t\left( Q\right) \widetilde{G}_{L,R}\left( \mathbf{q},\omega _{n}\right)
\label{system2debut} \\
-\sum_{\mathbf{q}^{\prime }}\widetilde{U}_{R,R}\left( \mathbf{q,q}^{\prime
}\right) \widetilde{G}_{R,R}\left( \mathbf{q}^{\prime },\omega _{n}\right)
-\sum_{\mathbf{q}^{\prime }}\widetilde{U}_{R,L}\left( \mathbf{q,q}^{\prime
}\right) \widetilde{G}_{L,R}\left( \mathbf{q}^{\prime },\omega _{n}\right)
=\delta _{\mathbf{q},0},  \notag
\end{gather}

\begin{gather}
\left[ i\omega _{n}-\frac{1}{\hslash }\left( E_{0}-\mu \right) \right] 
\widetilde{G}_{R,L}\left( \mathbf{q},\omega _{n}\right) +\frac{1}{\hslash }%
t\left( Q\right) \widetilde{G}_{L,L}\left( \mathbf{q},\omega _{n}\right) \\
-\sum_{\mathbf{q}^{\prime }}\widetilde{U}_{R,R}\left( \mathbf{q,q}^{\prime
}\right) \widetilde{G}_{R,L}\left( \mathbf{q}^{\prime },\omega _{n}\right)
-\sum_{\mathbf{q}^{\prime }}\widetilde{U}_{R,L}\left( \mathbf{q,q}^{\prime
}\right) \widetilde{G}_{L,L}\left( \mathbf{q}^{\prime },\omega _{n}\right)
=0,  \notag
\end{gather}%
\begin{gather}
\left[ i\omega _{n}-\frac{1}{\hslash }\left( E_{0}-\mu \right) \right] 
\widetilde{G}_{L,R}\left( \mathbf{q},\omega _{n}\right) +\frac{1}{\hslash }%
t\left( Q\right) \widetilde{G}_{R,R}\left( \mathbf{q},\omega _{n}\right) \\
-\sum_{\mathbf{q}^{\prime }}\widetilde{U}_{L,R}\left( \mathbf{q,q}^{\prime
}\right) \widetilde{G}_{R,R}\left( \mathbf{q}^{\prime },\omega _{n}\right)
-\sum_{\mathbf{q}^{\prime }}\widetilde{U}_{L,L}\left( \mathbf{q,q}^{\prime
}\right) \widetilde{G}_{L,R}\left( \mathbf{q}^{\prime },\omega _{n}\right)
=0,  \notag
\end{gather}%
\begin{gather}
\left[ i\omega _{n}-\frac{1}{\hslash }\left( E_{0}-\mu \right) \right] 
\widetilde{G}_{L,L}\left( \mathbf{q},\omega _{n}\right) +\frac{1}{\hslash }%
t\left( Q\right) \widetilde{G}_{R,L}\left( \mathbf{q},\omega _{n}\right)
\label{system2fin} \\
-\sum_{\mathbf{q}^{\prime }}\widetilde{U}_{L,R}\left( \mathbf{q,q}^{\prime
}\right) \widetilde{G}_{R,L}\left( \mathbf{q}^{\prime },\omega _{n}\right)
-\sum_{\mathbf{q}^{\prime }}\widetilde{U}_{L,L}\left( \mathbf{q,q}^{\prime
}\right) \widetilde{G}_{L,L}\left( \mathbf{q}^{\prime },\omega _{n}\right)
=\delta _{\mathbf{q},0}.  \notag
\end{gather}%
The four mean-field potentials, in this frame, are given by%
\begin{equation}
\widetilde{U}_{R,R}\left( \mathbf{q,q}^{\prime }\right) =\gamma ^{\ast
}\left( \mathbf{q}\right) U_{R,R}\left( \mathbf{q,q}^{\prime }\right) \gamma
\left( \mathbf{q}^{\prime }\right) ,
\end{equation}%
\begin{equation}
\widetilde{U}_{L,L}\left( \mathbf{q,q}^{\prime }\right) =\gamma \left( 
\mathbf{q}\right) U_{L,L}\left( \mathbf{q,q}^{\prime }\right) \gamma ^{\ast
}\left( \mathbf{q}^{\prime }\right) ,
\end{equation}%
\begin{eqnarray}
\widetilde{U}_{R,L}\left( \mathbf{q,q}^{\prime }\right) &=&U_{R,L}\left( 
\mathbf{q+Q,q}^{\prime }\right) \gamma ^{\ast }\left( \mathbf{q}^{\prime
}\right) , \\
\widetilde{U}_{L,R}\left( \mathbf{q,q}^{\prime }\right) &=&U_{L,R}\left( 
\mathbf{q-Q,q}^{\prime }\right) \gamma \left( \mathbf{q}^{\prime }\right) .
\end{eqnarray}%
Finally, the Hartree-Fock energy per particle is given by%
\begin{eqnarray}
E_{CSP} &=&-\frac{2}{\nu }\sum_{\alpha }t_{N}\left( \mathbf{Q}\right) \func{%
Re}\left[ \left\langle \widetilde{\rho }_{R,L}\left( 0\right) \right\rangle %
\right]  \label{energycsp} \\
&&+\frac{1}{2\nu }\left( \frac{e^{2}}{\kappa l_{\bot }}\right) \sum_{\mathbf{%
q}\neq 0}H\left( \mathbf{q}\right) \left[ \left| \left\langle \widetilde{%
\rho }_{R,R}\left( \mathbf{q}\right) \right\rangle \right| ^{2}+\left|
\left\langle \widetilde{\rho }_{L,L}\left( \mathbf{q}\right) \right\rangle
\right| ^{2}\right]  \notag \\
&&-\frac{1}{2\nu }\left( \frac{e^{2}}{\kappa l_{\bot }}\right) \sum_{\mathbf{%
q}}X\left( \mathbf{q}\right) \left[ \left| \left\langle \widetilde{\rho }%
_{R,R}\left( \mathbf{q}\right) \right\rangle \right| ^{2}+\left|
\left\langle \widetilde{\rho }_{L,L}\left( \mathbf{q}\right) \right\rangle
\right| ^{2}\right]  \notag \\
&&+\frac{1}{\nu }\left( \frac{e^{2}}{\kappa l_{\bot }}\right) \sum_{\mathbf{q%
}\neq 0}\widetilde{H}\left( \mathbf{q}\right) \left[ \gamma ^{\ast }\left( 
\mathbf{q}\right) \right] ^{2}\left[ \left\langle \widetilde{\rho }%
_{R,R}\left( -\mathbf{q}\right) \right\rangle \left\langle \widetilde{\rho }%
_{L,L}\left( \mathbf{q}\right) \right\rangle \right]  \notag \\
&&-\frac{1}{\nu }\left( \frac{e^{2}}{\kappa l_{\bot }}\right) \sum_{\mathbf{q%
}}\widetilde{X}\left( \mathbf{q+Q}\right) \left| \left\langle \widetilde{%
\rho }_{L,R}\left( \mathbf{q}\right) \right\rangle \right| ^{2}.  \notag
\end{eqnarray}

At fixed $d/l_{\bot }$ and $B_{\Vert }$, the energy of the CSP is obtained
by minimizing the Hartree-Fock energy of Eq.~(\ref{energycsp}) with respect
to $a$, the separation between the stripes in a given well (which may be
written in unitless form as $\xi =a/l_{\bot }$) and with respect to the
orientation $\alpha $ of the stripes with the $x$ axis. When $\alpha \neq 0$%
, the commensurate rotation of the pseudospins will induce a modulation in
the total density. As long as the rotation remains commensurate and has a
single wavevector, however, the induced density will have the same
periodicity as that of the stripes. This can be seen as follows. The
relation between the topological charge and the pseudospin texture is given
by 
\begin{equation}
\delta \left\langle n\left( \mathbf{r}\right) \right\rangle =\frac{1}{8\pi }%
\varepsilon _{abc}S_{a}\left( \mathbf{r}\right) \varepsilon _{ij}\partial
_{i}S_{b}\left( \mathbf{r}\right) \partial _{j}S_{c}\left( \mathbf{r}\right)
,  \label{topolo}
\end{equation}%
where $\varepsilon _{ij}$ and $\varepsilon _{abc}$ are antisymmetric tensors
and $\mathbf{S}\left( \mathbf{r}\right) $ is a classical field with unit
modulus representing the pseudospins. If we write a general solution as

\begin{eqnarray}
S_{x}\left( \mathbf{r}\right) &=&\sin \left[ \theta \left( \mathbf{r}\right) %
\right] \cos \left[ \varphi \left( \mathbf{r}\right) \right] , \\
S_{y}\left( \mathbf{r}\right) &=&\sin \left[ \theta \left( \mathbf{r}\right) %
\right] \sin \left[ \varphi \left( \mathbf{r}\right) \right] , \\
S_{z}\left( \mathbf{r}\right) &=&\cos \left[ \theta \left( \mathbf{r}\right) %
\right] ,
\end{eqnarray}%
then the induced density takes the simple form 
\begin{equation}
\delta n\left( \mathbf{r}\right) =\frac{1}{4\pi }\sin \left[ \theta \left( 
\mathbf{r}\right) \right] \left[ \nabla \varphi \left( \mathbf{r}\right)
\times \nabla \theta \left( \mathbf{r}\right) \right] \cdot \widehat{\mathbf{%
z}}.
\end{equation}%
Writing a general solution for $\varphi \left( \mathbf{r}\right) $ in the
form 
\begin{equation}
\varphi \left( \mathbf{r}\right) =\widetilde{\varphi }\left( \mathbf{r}%
\right) +\mathbf{Q}\cdot \mathbf{r},
\end{equation}%
the induced density can be written as 
\begin{equation}
\delta n\left( \mathbf{r}\right) =\frac{1}{4\pi }\sin \left[ \theta \left( 
\mathbf{r}\right) \right] \left[ \nabla \widetilde{\varphi }\left( \mathbf{r}%
\right) \times \nabla \theta \left( \mathbf{r}\right) +\mathbf{Q}\times
\nabla \theta \left( \mathbf{r}\right) \right] \cdot \widehat{\mathbf{z}}.
\end{equation}%
In the CSP and ISP, $\widetilde{\varphi }\left( \mathbf{r}\right) =0$ and so%
\begin{equation}
\delta n\left( \mathbf{r}\right) =\frac{1}{4\pi }\sin \left[ \theta \left( 
\mathbf{r}\right) \right] \left[ \mathbf{Q}\times \nabla \theta \left( 
\mathbf{r}\right) \right] \cdot \widehat{\mathbf{z}}.
\end{equation}%
There is no induced charge for stripes aligned with the parallel field [i.e. 
$\mathbf{Q\Vert }\nabla \theta \left( \mathbf{r}\right) $]. For other
orientations of the stripes, the induce charge has the same periodicity as
that of the angle $\theta \left( \mathbf{r}\right) $ that defines the
staggered magnetization in the $xz$ plane.

In our Hartree-Fock calculation, we find that the most favorable
configuration is that of the stripes aligned with the parallel magnetic
field. The same conclusion was also reported previously by Demler et al.\cite%
{demler}. Thus, we will choose $\alpha =0$ in the rest of this paper.

\subsection{Incommensurate Stripe Phase (ISP)}

In analogy with the IL state, we define an incommensurate stripe phase (ISP)
by the absence of rotation of the spins in the $xy$ plane. This state is
most naturally described by the original order parameters $\left\langle \rho
_{ij}(\mathbf{{q})}\right\rangle $ which, together with $G_{L,R}\left( 
\mathbf{q},\omega _{n}\right) $, are non-vanishing for $\mathbf{q}\in \{%
\mathbf{G}\}$. In particular, $G_{L,R}\left( \mathbf{q-Q},\omega _{n}\right)
=0$ for any $\mathbf{q}\in \{\mathbf{G}\}$ so that the tunneling term in
Eqs. (\ref{system1debut1})-(\ref{system1fin}) is zero. The order parameters
of the ISP are thus found by solving directly Eqs. (\ref{system1debut1})-(%
\ref{system1fin}) with $t\left( \mathbf{q}\right) =0$ (or, alternatively,
Eqs. (\ref{system2debut})-(\ref{system2fin}) with $t_{N}\left( Q\right) =0$ 
\textbf{and} $\mathbf{Q}=0$). While this is not an exact solution to the
Hartree-Fock equations, what is ignored are the remnants of the solitons of
the SLS which are relatively dense and leave only a very weak modulation of
the pseudospin, except quite close to the transition out of the CSP state %
\cite{cotesls}. Setting $t\left( \mathbf{q}\right) =0$ effectively ignores
this weak modulation, and introduces a great simplification in the numerics.
The resulting energy one finds for the ISP does not depend on the parallel
magnetic field, just as for the IL state. We note that, because of the
absence of tunneling in the Hartree-Fock equations, our approximate form of
the ISP is equivalent to the coherent stripe state studied in Refs.~%
\onlinecite{breydoublestripes} and \onlinecite{cotedoublestripes1}.

\section{Phase Diagram in Landau Level $N=2$}

In the Hartree-Fock approximation, the transition between the CL to the IL
states occurs when $E_{CL}=E_{IL}$ or, equivalently, when

\begin{equation}
\frac{\frac{1}{4}\left[ \widetilde{X}\left( 0\right) -\widetilde{X}\left( 
\mathbf{Q}\right) \right] e^{Q^{2}\ell _{\bot }^{2}/4}}{\left|
L_{n}^{0}\left( \frac{Q^{2}\ell _{\bot }^{2}}{2}\right) \right| }=\frac{t_{0}%
}{\left( e^{2}/\kappa \ell \right) }.
\end{equation}%
In Fig.~1, we show the contour lines in the $d-B_{_{\Vert }}$ plane
separating the CL from the IL for several values of the tunneling parameter.
(We assume a typical density of $n=1\times 10^{11}$ cm$^{-2}$ in our
calculation and in all the graphs presented in this paper). We remark that
the oscillations of $t_{N}\left( Q\right) $ lead to a complex sequence of
transitions between the two liquid states with two reentrant CL phases for $%
N=2$. We do not analyse these transitions in further detail, however,
because at a value of $d>d_{c}\left( B_{_{\Vert }}\right) $, the liquid
states are higher in energy than the stripe states as indicated in Fig.~1,
and the value of $d_{c}\left( B_{_{\Vert }}\right) $ is lower than the value
of $d$ at which the reentrant CL states appear. It is nevertheless an
interesting possibility that these reentrant transitions might occur when
Landau level mixing or finite well thickness effects are included in the
calculation.

When stripe phases are taken into account, the phase diagram becomes even
more complex because the oscillations of $t_{N}\left( Q\right) $ also
produce reentrant CSP and ISP\ states. Figure 5 shows the variation of the
energy of the four states with respect to $d$ for some typical cases. The
interchange of the ISP and CSP is clearly visible in Fig.~5(a). The line
called $d_{c}\left( B_{_{\Vert }}\right) $ in Fig.~1 indicates the
transition from one of the liquid states to one of the stripe states (either
CSP or ISP). At small value of $t_{0}$, $d_{c}$ is almost independent of $%
B_{_{\Vert }}$. At larger values of $d$ or $B_{_{\Vert }}$ where $%
t_{N}\left( Q\right) \rightarrow 0$, the amplitude of the pseudo-magnetic
field goes to zero and there is no gain in tunneling energy for the
pseudospins to rotate. On the contrary, there is only an exchange energy
cost. It is thus clear that the ground state must be the ISP at large $d$ or 
$B_{_{\Vert }}$ and the CSP only exists in a limited parameter region. Note
however that, at large value of $d$, the LCR's are very narrow and the cost
of the commensurate rotation in exchange energy is very small. The
difference between the CSP and ISP energies in this limit is close to our
numerical accuracy as can be seen in for large $d$ in Fig.~5(b) and large $%
B_{_{\Vert }}$ in Fig.~6(a).

\begin{figure}[tbp]
\includegraphics[width=16cm]{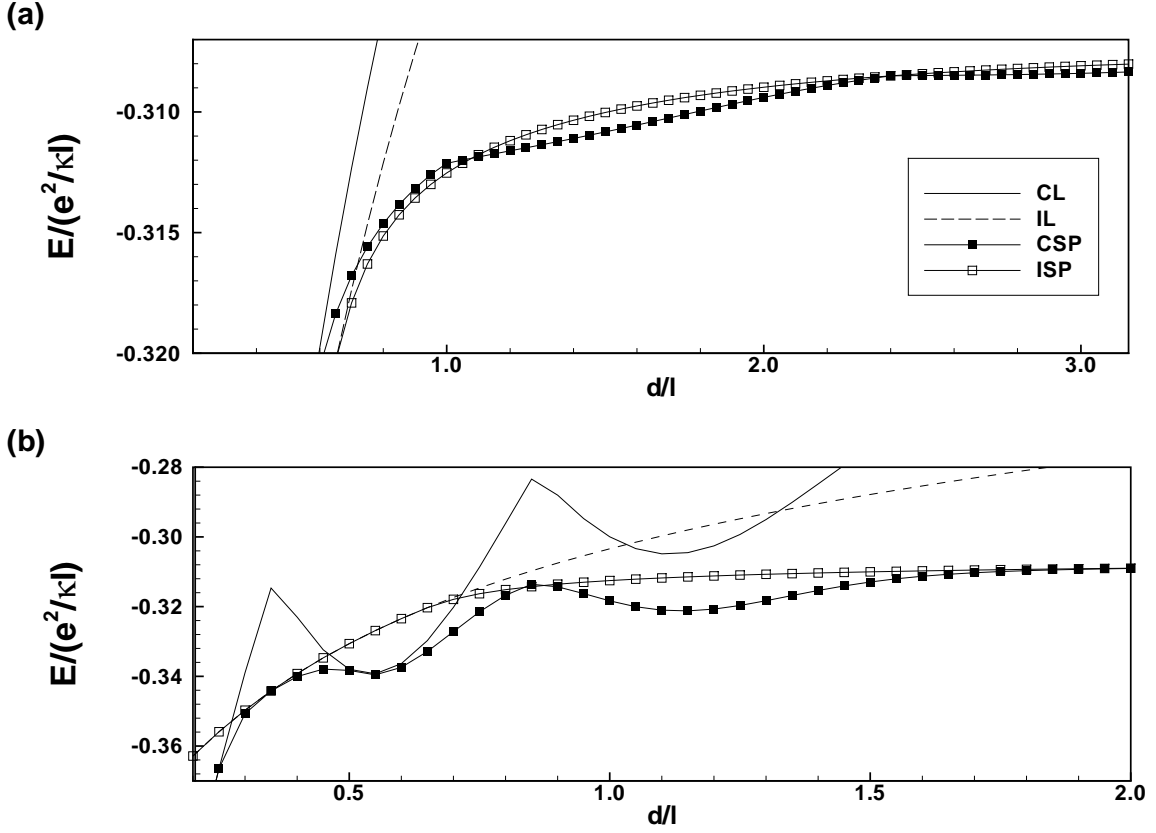}
\caption{Energy per electron in the liquid and stripe phases as a function
of the separation between the wells for $N=2$ and (a) $t_{0}/\left( e^{2}/%
\protect\kappa \ell \right) =0.01,B_{\Vert }=0.5$ T and (b) $t_{0}/\left(
e^{2}/\protect\kappa \ell \right) =0.1,$ $B_{\Vert }=1.4$ T.}
\label{fig5}
\end{figure}

~From an experimental point of view, it is more interesting to study the
behavior of the energy of the four states with respect to $B_{_{\Vert }}$ at
a fixed value of $d$. Referring to Fig.~1(a), we choose, for $t_{0}=0.01$, $%
d/\ell _{\bot }=1.4$, a value greater than $d_{c}\left( B_{_{\Vert }}\right) 
$ so that the ground state is one of the stripe states. We plot, in Fig.~6:
(a) the energy of the CSP and ISP, (b) the variation in $t_{N}\left(
Q\right) ,$ and (c) the corresponding variation of the order parameters $%
\left\langle \widetilde{\rho }_{R,L}\left( 0\right) \right\rangle _{CSP}$
and $\left\langle \rho _{R,L}\left( 0\right) \right\rangle _{ISP}$ . (We
remark that, in the CL, $\left\langle \widetilde{\rho }_{R,L}\left( 0\right)
\right\rangle =0.5$ and $\left\langle \rho _{R,L}\left( 0\right)
\right\rangle =0$ while in the IL, $\left\langle \widetilde{\rho }%
_{R,L}\left( 0\right) \right\rangle =0$ and $\left\langle \rho _{R,L}\left(
0\right) \right\rangle =0.5$). In the specific case studied in Fig.~6, the
ground state goes through the sequence CSP-ISP-CSP-ISP-CSP-ISP when $%
B_{_{\Vert }}$ is increased. A change in the sign of $t_{N}\left( Q\right) $
leads to a cusp in the curve of the energy of the CSP and to a corresponding
change in sign in $\left\langle \widetilde{\rho }_{R,L}\left( 0\right)
\right\rangle $. With our choice of phase, the pseudospins are oriented
along $x$ in the rotating frame for the CSP. When $t_{N}\left( Q\right) $
changes sign, the corresponding effective magnetic field is reversed and the
pseudospins then point along $-x$. In computing Fig.~6, we have included
only the CSP and ISP. But, we know from the study of the effect of $%
B_{_{\Vert }}$ on the ground state at $\nu =1$ that the commensurate state
does not go directly into the incommensurate state. At a certain $B_{_{\Vert
}}=B_{_{\Vert },CL-SLS}$, which is smaller than the value $B_{_{\Vert
}}=B_{_{\Vert },CL-IL}$ at which the $CL-IL$ transition occurs, the
commensurate state becomes unstable towards the formation of solitons. For $%
B_{\Vert }>B_{_{\Vert },CL-SLS}$, the ground state contains a finite density
of these defects. They form a soliton lattice whose lattice spacing depends
on $B_{_{\Vert }}$. In analogy with the case $\nu =1$, we expect that the
same scenario will repeat itself here whenever the CSP approaches the ISP
from below or above. A precursor of this instability is seen in the
softening of the low-energy modes of the CSP as described below \cite{com}.

\begin{figure}[tbp]
\includegraphics[width=16cm]{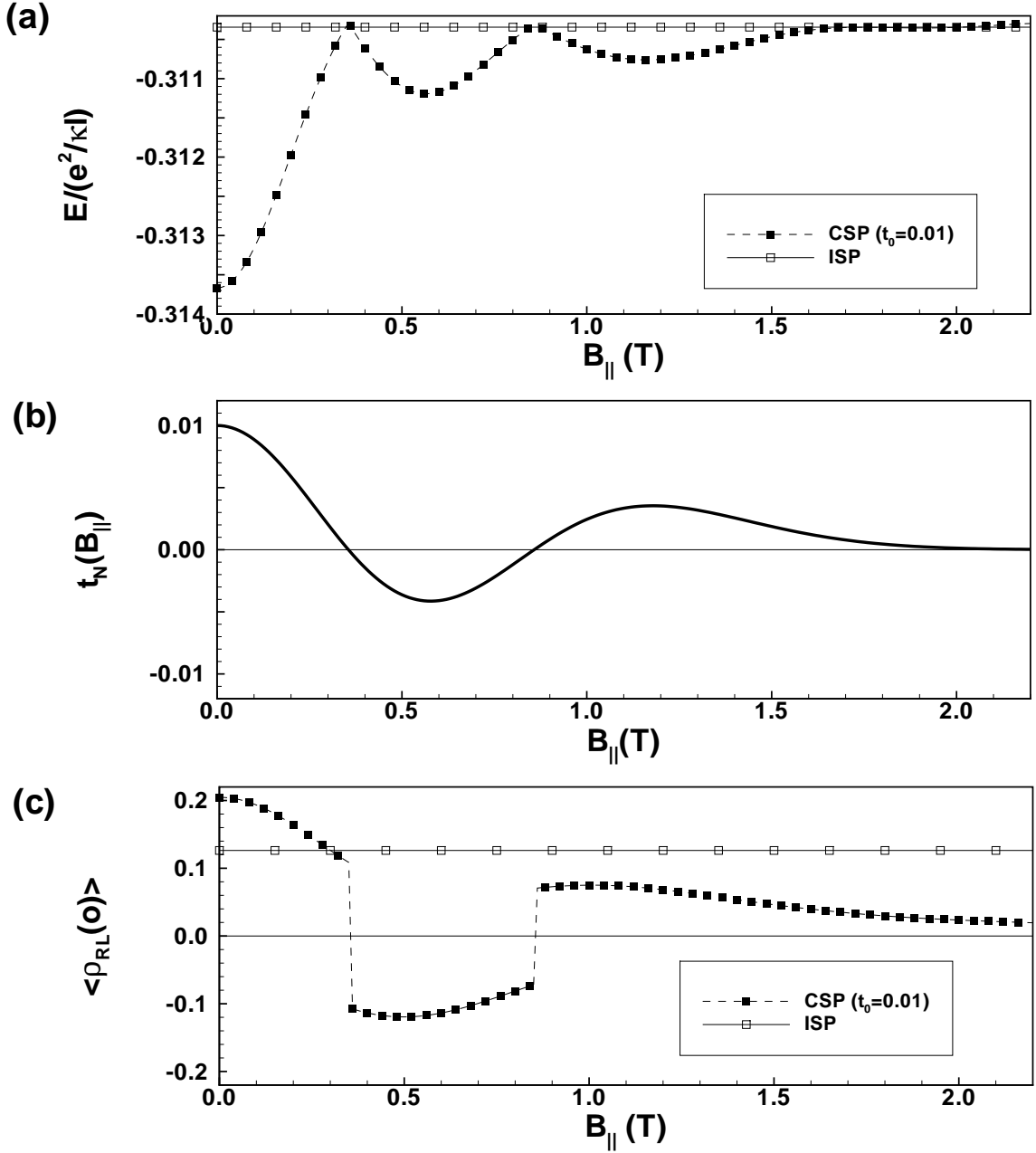}
\caption{Study of the CSP-ISP\ transition at $N=2$ for $t_{0}=0.01$ and $%
d/\ell _{\bot }=1.4$ showing (a) the variation of the energy of these two
phases with $B_{\Vert }$, (b) the variation of the tunneling amplitude $%
t_{N}\left( Q\right) $ with $B_{\Vert }$ and (c) the variations of the order
parameters $\left\langle \widetilde{\protect\rho }_{R,L}\left( 0\right)
\right\rangle _{CSP}$ and $\left\langle \protect\rho _{R,L}\left( 0\right)
\right\rangle _{ISP}$ with $B_{\Vert }$. }
\label{fig6}
\end{figure}

There are many possible transitions between the four states kept in our
analysis. In Fig.~1, we indicate all transitions between the CL and IL
states, but not all transitions into commensurate or incommensurate stripe
phases that occur for $d>d_{c}$. In general, we expect transitions between
the CSP and the ISP whenever the tunneling matrix element becomes small, as
occurs in the liquid states. Thus, just as we see $d_{c}$ occurs near the
first CL-IL transition in Fig.~1, we expect higher order CSP-ISP transitions
near what are shown as the reentrant CL-IL transitions. Clearly, this system
supports a remarkably rich phase diagram.

How might one detect these phase transitions in experiment? One possibility,
as we discuss below, is to study the behavior of the tunneling current with
the parallel magnetic field. As we shall see this affords a clear signature
of the transition between liquid and stripe phases. To compute the tunneling
current, we need to know the collective excitations and response functions
related to the density and pseudospin orders. We address this question in
the next section.

\section{Response Functions in the Time-Dependent Hartree-Fock Approximation}

As mentioned above, the collective modes are important in determining the
tunneling current in the double layer system in response to an interlayer
bias potential. Their dispersions can also be measured experimentally via
inelastic light scattering. To derive them, we compute the retarded density
and pseudospin response functions, and track their poles. The response
functions are obtained by analytical continuation of the two-particle
Matsubara Green's functions 
\begin{equation}
\chi _{i,j,k,l}\left( \mathbf{q,q}^{\prime };\tau \right) =-N_{\phi
}\left\langle T\rho _{i,j}\left( \mathbf{q,}\tau \right) \rho _{k,l}\left( -%
\mathbf{q}^{\prime },0\right) \right\rangle +N_{\phi }\left\langle \rho
_{i,j}\left( \mathbf{q}\right) \right\rangle \left\langle \rho _{k,l}\left( -%
\mathbf{q}^{\prime }\right) \right\rangle .  \label{responsefunctions}
\end{equation}%
The response functions $\chi _{i,j,k,l}\left( \mathbf{q,q}^{\prime };\tau
\right) $ are computed in the time-dependent Hartree-Fock approximation. The
calculation we do here is a generalization of that presented in Refs.~%
\onlinecite{cotedoublestripes1} and \onlinecite{cotesls}. To include the
parallel field and so the commensurate rotation, it is necessary to derive
the equation of motion for the matrix of response functions defined in the
rotating frame by%
\begin{equation}
\widetilde{P}\left( \mathbf{q},\mathbf{q}^{\prime };i\Omega _{n}\right)
\equiv \left( 
\begin{array}{cccc}
\gamma ^{\ast }\chi _{R,R,R,R}\left( \mathbf{q},\mathbf{q}^{\prime }\right)
\gamma & \gamma ^{\ast }\chi _{R,R,L,R}\left( \mathbf{q},\mathbf{q}^{\prime
}-\mathbf{Q}\right) & \gamma ^{\ast }\chi _{R,R,R,L}\left( \mathbf{q},%
\mathbf{q}^{\prime }+\mathbf{Q}\right) & \gamma ^{\ast }\chi
_{R,R,L,L}\left( \mathbf{q},\mathbf{q}^{\prime }\right) \gamma ^{\ast } \\ 
\chi _{R,L,R,R}\left( \mathbf{q-Q},\mathbf{q}^{\prime }\right) \gamma & \chi
_{R,L,L,R}\left( \mathbf{q-Q},\mathbf{q}^{\prime }-\mathbf{Q}\right) & \chi
_{R,L,R,L}\left( \mathbf{q-Q},\mathbf{q}^{\prime }+\mathbf{Q}\right) & \chi
_{R,L,L,L}\left( \mathbf{q-Q},\mathbf{q}^{\prime }\right) \gamma ^{\ast } \\ 
\chi _{L,R,R,R}\left( \mathbf{q+Q},\mathbf{q}^{\prime }\right) \gamma & \chi
_{L,R,L,R}\left( \mathbf{q+Q},\mathbf{q}^{\prime }-\mathbf{Q}\right) & \chi
_{L,R,R,L}\left( \mathbf{q+Q},\mathbf{q}^{\prime }+\mathbf{Q}\right) & \chi
_{L,R,L,L}\left( \mathbf{q+Q},\mathbf{q}^{\prime }\right) \gamma ^{\ast } \\ 
\gamma \chi _{L,L,R,R}\left( \mathbf{q},\mathbf{q}^{\prime }\right) \gamma & 
\gamma \chi _{L,L,L,R}\left( \mathbf{q},\mathbf{q}^{\prime }-\mathbf{Q}%
\right) & \gamma \chi _{L,L,R,L}\left( \mathbf{q},\mathbf{q}^{\prime }+%
\mathbf{Q}\right) & \gamma \chi _{L,L,L,L}\left( \mathbf{q},\mathbf{q}%
^{\prime }\right) \gamma ^{\ast }%
\end{array}%
\right)
\end{equation}%
where $\gamma $ on the left(right) of $\widetilde{P}\left( \mathbf{q},%
\mathbf{q}^{\prime };\Omega _{n}\right) $ stands for $\gamma \left( \mathbf{q%
}\right) $($\gamma \left( \mathbf{q}^{\prime }\right) $) and $\Omega _{n}$
is a Matsubara bosonic frequency. In the time-dependent Hartree-Fock
approximation (TDHFA), the summation of ladders and bubble diagrams gives%
\begin{equation}
\widetilde{P}\left( \mathbf{q},\mathbf{q}^{\prime };i\Omega _{n}\right) =%
\widetilde{P}^{0}\left( \mathbf{q},\mathbf{q}^{\prime };i\Omega _{n}\right) +%
\frac{1}{\hslash }\sum_{\mathbf{q}^{\prime \prime }}\widetilde{P}^{0}\left( 
\mathbf{q},\mathbf{q}^{\prime \prime };i\Omega _{n}\right) Y\left( \mathbf{q}%
^{\prime \prime }\right) \widetilde{P}\left( \mathbf{q}^{\prime \prime },%
\mathbf{q}^{\prime };i\Omega _{n}\right) .  \label{grpasum}
\end{equation}%
The vertex corrections involve the matrix of interactions%
\begin{equation}
Y\left( \mathbf{q}\right) =\left( \frac{e^{2}}{\kappa l_{\bot }}\right)
\left( 
\begin{array}{cccc}
H\left( \mathbf{q}\right) -X\left( \mathbf{q}\right) & 0 & 0 & \widetilde{H}%
\left( \mathbf{q}\right) \left[ \gamma ^{\ast }\left( \mathbf{q}\right) %
\right] ^{2} \\ 
0 & -\widetilde{X}_{-}\left( \mathbf{q}\right) & 0 & 0 \\ 
0 & 0 & -\widetilde{X}_{+}\left( \mathbf{q}\right) & 0 \\ 
\widetilde{H}\left( \mathbf{q}\right) \left[ \gamma \left( \mathbf{q}\right) %
\right] ^{2} & 0 & 0 & H\left( \mathbf{q}\right) -X\left( \mathbf{q}\right)%
\end{array}%
\right) ,
\end{equation}%
where we have defined%
\begin{equation}
\widetilde{X}_{\pm }\left( \mathbf{q}\right) =\widetilde{X}\left( \mathbf{q}%
\pm \mathbf{Q}\right) .
\end{equation}%
Using the Hartree-Fock Hamiltonian, $H_{HF}$, written in terms of the $%
\left\langle \rho _{i,j}\left( \mathbf{q}\right) \right\rangle $, we obtain
an equation of motion for the Hartree-Fock response functions $\widetilde{P}%
^{0}\left( \mathbf{q},\mathbf{q}^{\prime };i\Omega _{n}\right) $ using $%
\frac{\partial }{\partial \tau }\left( \ldots \right) =\left[ H_{HF}-\mu
N,\left( \ldots \right) \right] $. Combining with Eq.~(\ref{grpasum}), we
obtain after a lengthy calculation an equation of motion for the TDHFA
response functions than can be put in the simple matrix form 
\begin{equation}
\sum_{k}\sum_{\mathbf{q}^{\prime \prime }}\left[ i\hslash \Omega _{n}\delta
_{i,k}\delta _{\mathbf{q},\mathbf{q}^{\prime \prime }}-F_{i,k}\left( \mathbf{%
q},\mathbf{q}^{\prime \prime }\right) \right] \widetilde{P}_{k,j}\left( 
\mathbf{q}^{\prime \prime },\mathbf{q}^{\prime }\right) =D_{i,j}\left( 
\mathbf{q},\mathbf{q}^{\prime }\right) .  \label{simpler}
\end{equation}%
For completeness, we give the definitions of the matrices $F$ and $D$ in the
Appendix. Notice that the dynamical response can be computed from a
knowledge of the static order parameters $\left\{ \left\langle \rho
_{i,j}\left( \mathbf{q}\right) \right\rangle \right\} $ alone.

\subsection{Response Functions of the Commensurate Liquid}

For the CL, we can solve analytically Eq.~(\ref{simpler}) to get (after the
analytical continuation $i\Omega _{n}\rightarrow \omega +i\delta $ and with $%
\zeta =sign\left[ t_{N}\left( Q\right) \right] $)%
\begin{eqnarray}
&&\left( 
\begin{array}{cccc}
\chi _{R,R,R,R}\left( \mathbf{q},\mathbf{q}^{\prime }\right) & \chi
_{R,R,L,R}\left( \mathbf{q},\mathbf{q}^{\prime }-\mathbf{Q}\right) & \chi
_{R,R,R,L}\left( \mathbf{q},\mathbf{q}^{\prime }+\mathbf{Q}\right) & \chi
_{R,R,L,L}\left( \mathbf{q},\mathbf{q}^{\prime }\right) \\ 
\chi _{R,L,R,R}\left( \mathbf{q-Q},\mathbf{q}^{\prime }\right) & \chi
_{R,L,L,R}\left( \mathbf{q-Q},\mathbf{q}^{\prime }-\mathbf{Q}\right) & \chi
_{R,L,R,L}\left( \mathbf{q-Q},\mathbf{q}^{\prime }+\mathbf{Q}\right) & \chi
_{R,L,L,L}\left( \mathbf{q-Q},\mathbf{q}^{\prime }\right) \\ 
\chi _{L,R,R,R}\left( \mathbf{q+Q},\mathbf{q}^{\prime }\right) & \chi
_{L,R,L,R}\left( \mathbf{q+Q},\mathbf{q}^{\prime }-\mathbf{Q}\right) & \chi
_{L,R,R,L}\left( \mathbf{q+Q},\mathbf{q}^{\prime }+\mathbf{Q}\right) & \chi
_{L,R,L,L}\left( \mathbf{q+Q},\mathbf{q}^{\prime }\right) \\ 
\chi _{L,L,R,R}\left( \mathbf{q},\mathbf{q}^{\prime }\right) & \chi
_{L,L,L,R}\left( \mathbf{q},\mathbf{q}^{\prime }-\mathbf{Q}\right) & \chi
_{L,L,R,L}\left( \mathbf{q},\mathbf{q}^{\prime }+\mathbf{Q}\right) & \chi
_{L,L,L,L}\left( \mathbf{q},\mathbf{q}^{\prime }\right)%
\end{array}%
\right)  \label{chicl} \\
&=&\frac{\left( \zeta /2\right) \delta _{\mathbf{q},\mathbf{q}^{\prime }}}{%
\hslash \left[ \left( \omega +i\delta \right) ^{2}-\omega _{CL}^{2}\left( 
\mathbf{q}\right) \right] }\left( 
\begin{array}{cccc}
\left( b+c\right) & -\hslash \left( \omega +i\delta \right) \gamma & \hslash
\left( \omega +i\delta \right) \gamma & -\left( b+c\right) \gamma ^{2} \\ 
-\hslash \left( \omega +i\delta \right) \gamma ^{\ast } & 2\func{Re}\left(
a\right) & -2\func{Re}\left( a\right) & \hslash \left( \omega +i\delta
\right) \gamma \\ 
\hslash \left( \omega +i\delta \right) \gamma ^{\ast } & -2\func{Re}\left(
a\right) & 2\func{Re}\left( a\right) & -\hslash \left( \omega +i\delta
\right) \gamma \\ 
-\left( \gamma ^{\ast }\right) ^{2}\left( b+c\right) & \hslash \left( \omega
+i\delta \right) \gamma ^{\ast } & -\hslash \left( \omega +i\delta \right)
\gamma ^{\ast } & +\left( b+c\right)%
\end{array}%
\right) ,  \notag
\end{eqnarray}%
where 
\begin{eqnarray}
a\left( \mathbf{q}\right) &=&\frac{1}{2}\zeta \left[ 2\left| t_{N}\left(
Q\right) \right| +\left( \frac{e^{2}}{\kappa l_{\bot }}\right) \widetilde{X}%
\left( Q\right) +\left( \frac{e^{2}}{\kappa l_{\bot }}\right) \left( H\left(
q\right) -X\left( q\right) -\widetilde{H}\left( q\right) e^{-i\mathbf{q}%
\times \mathbf{Q}l_{\bot }^{2}}\right) \right] , \\
b\left( \mathbf{q}\right) &=&\frac{1}{2}\zeta \left[ 2\left| t_{N}\left(
Q\right) \right| +\left( \frac{e^{2}}{\kappa l_{\bot }}\right) \left[ 
\widetilde{X}\left( Q\right) -\widetilde{X}_{+}\left( q\right) \right] %
\right] , \\
c\left( \mathbf{q}\right) &=&\frac{1}{2}\zeta \left[ 2\left| t_{N}\left(
Q\right) \right| +\left( \frac{e^{2}}{\kappa l_{\bot }}\right) \left[ 
\widetilde{X}\left( Q\right) -\widetilde{X}_{-}\left( q\right) \right] %
\right] .
\end{eqnarray}%
There is a unique collective mode corresponding to an elliptical motion of
the pseudospins about the $x$ axis, in the $yz$ plane, with a frequency
given by%
\begin{equation}
\hslash \omega _{CL}\left( \mathbf{q}\right) =\sqrt{\left[ a\left( \mathbf{q}%
\right) +a^{\ast }\left( \mathbf{q}\right) \right] \left[ b\left( \mathbf{q}%
\right) +c\left( \mathbf{q}\right) \right] }  \label{omegacl}
\end{equation}%
(We remark that, for $\mathbf{q}=0$, $H\left( \mathbf{q}\right) -\widetilde{H%
}\left( \mathbf{q}\right) \cos \left[ \mathbf{q}\times \mathbf{Q}l_{\bot
}^{2}\right] \rightarrow -d/\ell _{\bot }$). Note $\omega _{CL}\left( 
\mathbf{q}\right) $ tends to a finite value for any $t_{N}(\mathbf{Q})>0$ in
the long-wavelength limit\cite{comlw}.

\subsection{Response Functions of the Incommensurate Liquid}

For the IL, the response functions are obtained by setting $t_{N}\left(
Q\right) =0$ and $Q=0$. We get%
\begin{eqnarray}
&&\left( 
\begin{array}{cccc}
\chi _{R,R,R,R}\left( \mathbf{q},\mathbf{q}^{\prime }\right) & \chi
_{R,R,L,R}\left( \mathbf{q},\mathbf{q}^{\prime }\right) & \chi
_{R,R,R,L}\left( \mathbf{q},\mathbf{q}^{\prime }\right) & \chi
_{R,R,L,L}\left( \mathbf{q},\mathbf{q}^{\prime }\right) \\ 
\chi _{R,L,R,R}\left( \mathbf{q},\mathbf{q}^{\prime }\right) & \chi
_{R,L,L,R}\left( \mathbf{q},\mathbf{q}^{\prime }\right) & \chi
_{R,L,R,L}\left( \mathbf{q},\mathbf{q}^{\prime }\right) & \chi
_{R,L,L,L}\left( \mathbf{q},\mathbf{q}^{\prime }\right) \\ 
\chi _{L,R,R,R}\left( \mathbf{q},\mathbf{q}^{\prime }\right) & \chi
_{L,R,L,R}\left( \mathbf{q},\mathbf{q}^{\prime }\right) & \chi
_{L,R,R,L}\left( \mathbf{q},\mathbf{q}^{\prime }\right) & \chi
_{L,R,L,L}\left( \mathbf{q},\mathbf{q}^{\prime }\right) \\ 
\chi _{L,L,R,R}\left( \mathbf{q},\mathbf{q}^{\prime }\right) & \chi
_{L,L,L,R}\left( \mathbf{q},\mathbf{q}^{\prime }\right) & \chi
_{L,L,R,L}\left( \mathbf{q},\mathbf{q}^{\prime }\right) & \chi
_{L,L,L,L}\left( \mathbf{q},\mathbf{q}^{\prime }\right)%
\end{array}%
\right)  \label{chiil} \\
&=&\frac{\left( 1/2\right) \delta _{\mathbf{q},\mathbf{q}^{\prime }}}{%
\hslash \left[ \left( \omega +i\delta \right) ^{2}-\omega _{IL}^{2}\left( 
\mathbf{q}\right) \right] }\left( 
\begin{array}{cccc}
2B & -\hslash \left( \omega +i\delta \right) & \hslash \left( \omega
+i\delta \right) & -2B \\ 
-\hslash \left( \omega +i\delta \right) & 2A & -2A & \hslash \left( \omega
+i\delta \right) \\ 
\hslash \left( \omega +i\delta \right) & -2A & 2A & -\hslash \left( \omega
+i\delta \right) \\ 
-2B & \hslash \left( \omega +i\delta \right) & -\hslash \left( \omega
+i\delta \right) & 2B%
\end{array}%
\right) ,  \notag
\end{eqnarray}%
where%
\begin{eqnarray}
A\left( \mathbf{q}\right) &=&\frac{1}{2}\left( \frac{e^{2}}{\kappa l_{\bot }}%
\right) \left[ H\left( \mathbf{q}\right) -\widetilde{H}\left( \mathbf{q}%
\right) +\widetilde{X}\left( 0\right) -X\left( \mathbf{q}\right) \right] ,
\label{ail} \\
B\left( \mathbf{q}\right) &=&\frac{1}{2}\left( \frac{e^{2}}{\kappa l_{\bot }}%
\right) \left[ \widetilde{X}\left( 0\right) -\widetilde{X}\left( \mathbf{q}%
\right) \right] .  \label{bil}
\end{eqnarray}%
The collective mode is gapless and given by%
\begin{equation}
\hslash \omega _{IL}\left( \mathbf{q}\right) =2\sqrt{A\left( \mathbf{q}%
\right) B\left( \mathbf{q}\right) }.  \label{omegail}
\end{equation}%
To linear order in $q$ (and for $d\neq 0$), this becomes 
\begin{equation}
\hslash \omega _{IL}\left( \mathbf{q}\right) \approx q\left( \frac{e^{2}}{%
\kappa l_{\bot }}\right) \sqrt{\left( \frac{-1}{2}\right) \left. \frac{%
\partial ^{2}\widetilde{X}\left( k\right) }{\partial k^{2}}\right|
_{k=0}\left( \widetilde{X}\left( 0\right) -X\left( 0\right) +\frac{d}{\ell
_{\bot }}\right) }
\end{equation}%
Thus for long wavelengths the dispersion is linear in $q$ and gapless even
if $t>0$.

\subsection{Response Functions in the Commensurate Stripe Phase}

We study the dispersion relation of the lower-energy collective modes of the
CSP by a numerical solution of the system of equations given by Eq.~(\ref%
{simpler}). In the absence of tunneling and parallel magnetic field, there
are two Goldstone modes \cite{cotedoublestripes1}. One is a phonon mode with
an in-phase motion of the density of the stripes in the two wells
accompanied by an $x-z$ motion of the pseudospins. The second is a
pseudospin wave mode with an out-of-phase motion of the densities and a
small rotation of the pseudospins about their equilibrium position. As
explained in Ref.~\onlinecite{cotedoublestripes1}, the low-energy modes can
be described by a pseudospin-wave theory where the effective periodicity is
set by the separation between neighboring linear coherent regions (LCR's)
which is half the separation between the stripes in a given layer. If we
take the the Brillouin zone edges (in the $x$ direction) to be $\pm 2\pi /a$
instead of $\pm \pi /a$, then the phonon and pseudospin wave mode are both
part of the same low-energy mode, the former dispersing from the zone center
and the latter from the zone edge. This is very clearly seen from Fig.~2(a)
and Fig.~2(b) where we show the full dispersion relation of these modes in
the ISP and in the CSP with $B_{\Vert }=0$T. In Fig.~2(b), we see that the
gap at the zone edge (and so in the pseudospin wave mode) is strongly
increased from its bare value $2t_{0}$ by self-energy and vertex
corrections. The phonon, by contrast, has vanishing energy at the zone
center. ~For $B_{\Vert }=0$, the phonon mode has $\omega \left( k_{\Vert
},k_{\bot }=0\right) \sim k_{\Vert }^{2}$ and $\omega \left( k_{\Vert
}=0,k_{\bot }\right) \sim k_{\bot }$ and is independent of $t_{0}$. For $%
B_{\Vert }=0$ and $t_{0}=0$, the pseudospin wave mode has $\omega \left(
k_{\Vert },k_{\bot }=0\right) \sim k_{\Vert }$, $\omega \left( k_{\Vert
}=0,k_{\bot }\right) \sim k_{\bot } $ (see Ref.~%
\onlinecite{cotedoublestripes1}).

In Fig.~6, we showed the behavior of the renormalized tunneling term, $%
t_{N}\left( Q\right) $, as $B_{_{\Vert }}$ is increased at $d/\ell _{\bot
}=1.4$ for $N=2$. We now look at the corresponding behavior of the
dispersion relation for the phonon and pseudospin wave modes. (As discussed
above, these two modes are both part of the same collective mode in an
extended Brillouin zone such as that used in Fig. 7). Figure 7 shows the
dispersion relations of these collective excitations as $B_{_{\Vert }}$ is
increased until slightly above the first value where $t_{N}\left( Q\right)
=0 $ where the system is in the ISP. The parallel field has an important
effect on the lowest-energy modes. The gap in the pseudospin mode (the part
of the dispersion near the zone edge in Fig. 7) is progressively closed as $%
t_{N}\left( Q\right) \rightarrow 0$. But, these modes do not become
completely dispersionless along $k_{\Vert }=0$. Before the CSP-ISP
transition, these low-energy modes become unstable at the zone center at
some value $B_{_{\Vert }}^{\ast }<B_{\Vert ,CSP-ISP}$ . The instability
persists in a small region around $B_{\Vert ,CSP-ISP}$. (For the case
represented in Fig.~6, we found that the first instability region is from $%
B_{\Vert }=0.34$ to $B_{\Vert }=0.40$). Above $B_{\Vert ,CSP-ISP}$, the
ground state is stable again and the gap increases and decreases again until
the next zero of $t_{N}\left( Q\right) =0$ where the same scenario is
repeated. Finally, just before the final transition from the CSP to the ISP,
i.e. at $B_{_{\Vert }}=1.6$ for the case represented in Fig.~6, the CSP
becomes unstable again.

\begin{figure}[tbp]
\includegraphics[width=8.25cm]{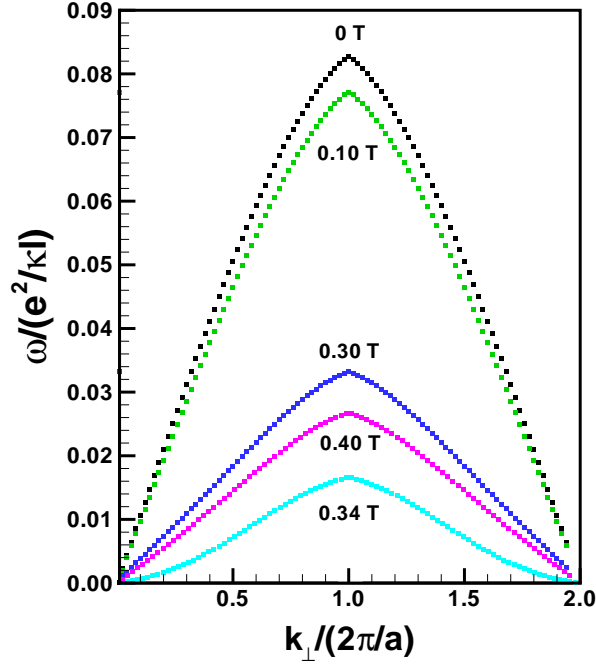}
\caption{Dispersion relation $\protect\omega \left( k_{\Vert }=0,k_{\bot
}\right) $ of the lowest-energy collective mode in the CSP for $%
N=2,t_{0}/\left( e^{2}/\protect\kappa \ell \right) =0.01,d/\ell _{\bot }=1.4$
and several values of the parallel magnetic field. The $\left[ 2\protect\pi %
/a,4\protect\pi /a\right] $ region can be translated left by $2\protect\pi %
/a $ to give the pseudospin wave dispersion. }
\label{fig7}
\end{figure}

In analogy with the case of filling factor $\nu =1$, we expect that these
instabilities are precursors of a transition from the CSP to a soliton
lattice state. For $\nu =1$, the instability occurs for $B_{_{\Vert }}^{\ast
}>B_{\Vert ,CL-SLS}$ and the transition from the commensurate liquid to the
SLS is first order. We do not know if this is the case for the CSP to ISP\
transition as we could not calculate $B_{\Vert ,CSP-SLS}$ with our
Hartree-Fock method. In any case, these instabilities need not be worried
about when considering the low energy spectrum: the introduction of solitons
into the groundstate will open gaps in the collective mode spectrum at
various values of $k_{\bot }$, but should not provide any new gapless modes
since translational symmetry has already been broken by the stripes.

Figure 8 shows the dispersion relation of the collective modes in the
direction of the stripes for several values of the parallel magnetic field.
It is clear that a change in $B_{_{\Vert }}$ has a profound impact on the
long-wavelength dispersion of the phase mode (lowest energy gapped mode at $%
k_{\Vert }=0$) as well as on the roton minimum. In constrast, the
higher-energy modes are much less affected by the parallel field. (This is
also true for dispersion perpendicular to the stripes). The phase mode
becomes nearly linear in $k_{\Vert }$ at $k_{\bot }=0$ near the zeros of $%
t_{N}\left( Q\right) $. The long-wavelength phonon dispersion changes slowly
with $B_{_{\Vert }}$. At small $B_{_{\Vert }}$, we find $\omega \left(
k_{\Vert }\rightarrow 0,k_{\bot }=0\right) \sim k_{\Vert }^{2}$ while at $%
B_{_{\Vert }}=0.8$ T (and for $d/\ell _{\bot }=1.4$), the dispersion crosses
over from quadratic to linear in $k_{\Vert }$ at a relatively small value of 
$k_{\Vert }$ [see Fig.~8(d)].

\begin{figure}[tbp]
\includegraphics[width=16.5cm]{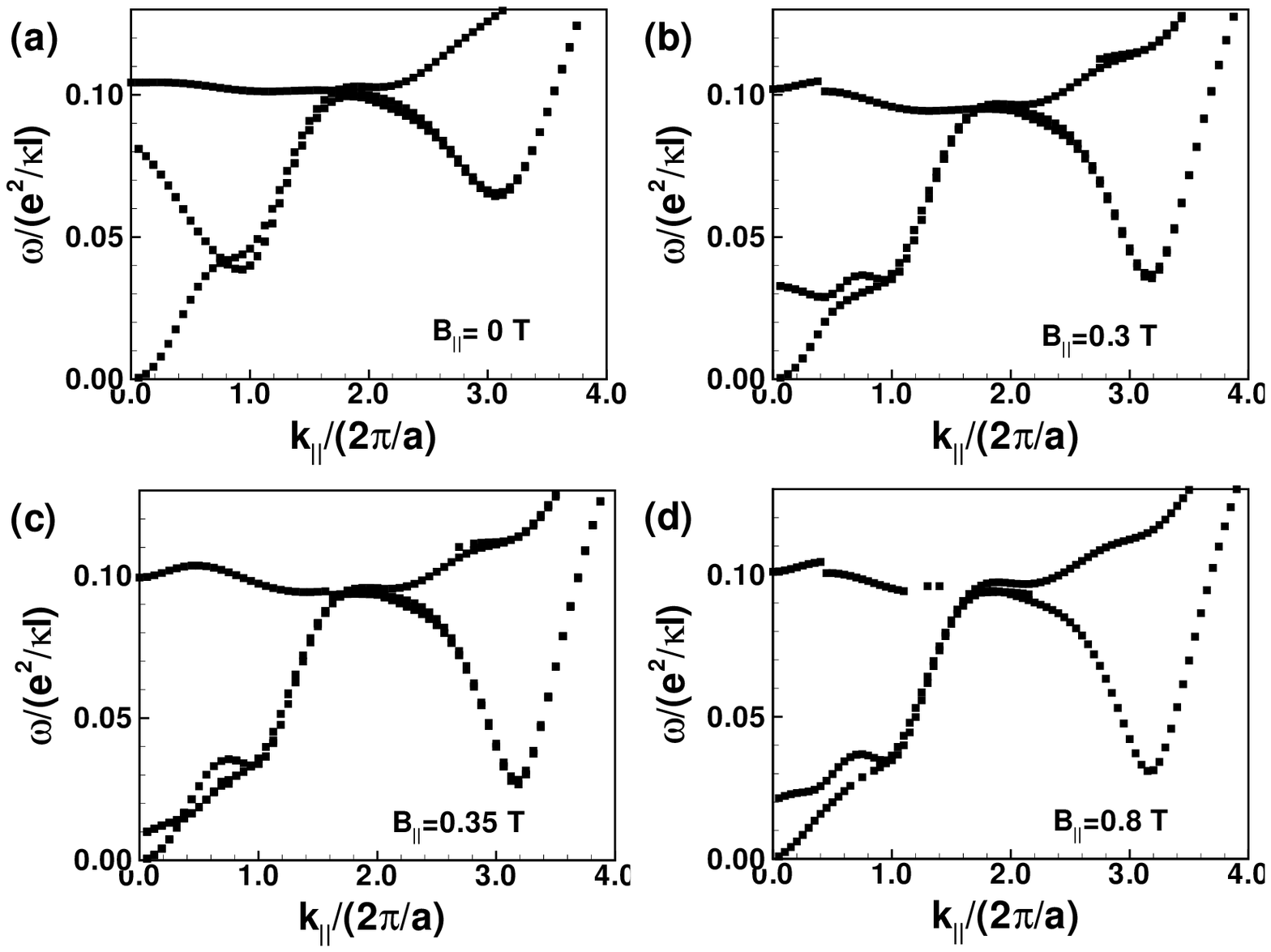}
\caption{Evolution of the dispersion relations $\protect\omega \left(
k_{\Vert },k_{\bot }=0\right) $ of the lowest-energy collective modes of the
CSP as a function of the parallel magnetic field for $N=2,t_{0}/\left( e^{2}/%
\protect\kappa \ell \right) =0.01$ and $d/\ell _{\bot }=1.4.$}
\label{fig8}
\end{figure}

As the separation between the wells increases, both the CSP and ISP\ are
unstable and the true ground state is that of a modulated stripe state
described in Ref.~\onlinecite{cotedoublestripes1}. For $t_{0}/\left(
e^{2}/\kappa \ell \right) =0.01$, we find that the critical spacing is $%
d/\ell _{\bot }\approx 2.0.$

\section{Tunneling response in the commensurate stripe state}

In this section we compute the interlayer tunneling $I-V$ to second order in
the tunneling matrix element $t_{0}$. Since $t_{0}$ is assumed to be small,
the commensurate states can only exist in a very narrow range of parameters $%
d$ and $B_{_{\Vert }}$, so that they may effectively be ignored. We are thus
interested in the IL-ISP transition, and whether there are distinguishable
signatures of these two phases in the tunneling $I-V$. Recent work \cite%
{spielman,tunneltheory} has demonstrated that a peak in the tunneling $I-V$
as a function of $B_{_{\Vert }}$ maps out the Goldstone mode associated with
spontaneous interlayer coherence in the IL. As we now show, this result
generalizes to the ISP, and the behavior of the peak with $B_{_{\Vert }}$
clearly distinguishes the two phases.

Our formalism closely follows that of Mahan \cite{mahan}. The tunneling
current is defined by 
\begin{equation}
I\left( t\right) =-e\frac{dN_{L}}{dt}=+e\frac{dN_{R}}{dt},
\end{equation}%
\newline
where $N_{L(R)}=\sum_{X}$ $c_{X,L(R)}^{\dagger }c_{X,L(R)}=N_{\phi }\rho
_{L(R),L(R)}\left( \mathbf{q}=0\right) $ is the number of electrons in the
left (right) well. $I\left( t\right) $ represents the interlayer current in
the presence of a bias voltage while the tunneling operator connecting the
two layers is adiabatically switched on, reaching its full value at time $%
t=0 $. The complete Hamiltonian of the system, including a bias between the
wells, is written as%
\begin{equation}
H=H^{\prime }+H_{T}=H_{0}-\mu _{L}N_{L}-\mu _{R}N_{R}+H_{T},
\end{equation}%
where $eV=\mu _{L}-\mu _{R}$ is the potential bias and $H_{T}$ is the
tunneling hamiltonian given in Eq.~(\ref{tunnel}). Written in terms of the $%
\rho $ operators, it takes the form ($\mathbf{Q=}Q\widehat{x}$) 
\begin{equation}
H_{T}=-N_{\phi }t_{N}\left( Q\right) \left[ \rho _{R,L}\left( -\mathbf{Q}%
\right) +\rho _{L,R}\left( \mathbf{Q}\right) \right] .
\end{equation}%
The term $H_{0}$ in $H$ contains the Coulomb interaction and kinetic
energies. Since $H_{T}$ is the only term that can change particle numbers in
one well, we have 
\begin{equation}
\frac{dN_{R}}{dt}=i\left[ H_{T},N_{R}\right] =\frac{i}{\hslash }N_{\phi
}t_{N}\left( Q\right) \left[ \rho _{R,L}\left( -\mathbf{Q}\right) -\rho
_{L,R}\left( \mathbf{Q}\right) \right] .
\end{equation}%
In all four states considered in this paper, $\left\langle \rho _{R,L}\left(
-\mathbf{Q}\right) \right\rangle =\left\langle \rho _{L,R}\left( \mathbf{Q}%
\right) \right\rangle $ so that there is no current in the ground state.

Using standard linear response theory \cite{mahan}, the current is given by%
\begin{equation}
\left\langle I\left( t\right) \right\rangle =\frac{i}{\hslash }\int_{-\infty
}^{t}dt^{\prime }\left\langle \left[ \widetilde{H}_{T}\left( t^{\prime
}\right) ,\widetilde{I}\left( t\right) \right] \right\rangle _{H^{\prime }},
\end{equation}%
where the time dependence of the operators is given by $\widetilde{I}\left(
t\right) =e^{iH^{\prime }t/\hslash }I\left( 0\right) e^{-iH^{\prime
}t/\hslash }$. After some algebra, we find the familiar result 
\begin{eqnarray}
I\left( t\right) &=&-2\frac{e}{\hslash ^{2}}N_{\phi }\left| t_{N}\left(
Q\right) \right| ^{2}\func{Im}\left[ \chi _{R,L,L,R}\left( \mathbf{-Q},-%
\mathbf{Q,}\hslash \omega =eV\right) \right]  \label{oldt} \\
&&-2\frac{e}{\hslash ^{2}}N_{\phi }\left| t_{N}\left( Q\right) \right| \func{%
Im}\left[ e^{-2ieVt}\chi _{R,L,R,L}\left( \mathbf{-Q},+\mathbf{Q,}\hslash
\omega \mathbf{=}eV\right) \right] .  \notag
\end{eqnarray}%
The response functions in Eq.~(\ref{oldt}) must be evaluated in a ground
state where tunneling and so the parallel magnetic field are absent. Thus,
only the response functions of the IL or ISP are relevant in this approach
to the tunneling $I-V$. In the liquid state where we have the analytical
result given by Eq.~(\ref{chiil}), the second term in Eq.~(\ref{oldt}) does
not contribute to the tunneling current and we find we the tunneling current
of the IL phase for small $t_{0}$, 
\begin{equation}
I_{IL}\left( t\right) =\frac{e}{2\hslash }N_{\phi }\left| t_{N}\left(
Q\right) \right| ^{2}\pi \sqrt{\frac{A\left( Q\right) }{B\left( Q\right) }}%
\delta \left( eV-\hslash \omega _{IL}\left( Q\right) \right) ,
\label{tunnel1}
\end{equation}%
where $A\left( Q\right) $ and $B\left( Q\right) $ are given by Eqs. (\ref%
{ail}) and (\ref{bil}) above and $\omega _{IL}\left( Q\right) $ is the
Goldstone mode of the IL state whose dispersion is given by Eq.~(\ref%
{omegail}). This result is identical to that of Ref.~%
\onlinecite{tunneltheory}, and as pointed out there the peak in the
tunneling current can be used to follow the dispersion relation of the
Goldstone mode. In the experiment of Spielman et al. \cite{spielman}, the
tunneling amplitude is extremely small and the expression above can be
applied provided one introduces a phenomenological broadening due to
disorder and thermal fluctuations. This can be achieved rather naturally by
substituting a Lorentzian for the delta function in Eq.~(\ref{tunnel1}),
although the microscopic mechanism leading to this broadening is currently
not understood.

Whereas in the IL state, the tunneling current peaks near zero bias only for 
$B_{_{\Vert }}=0$, in the ISP the Goldstone mode folds back upon itself,
continuously deforming into the phonon mode as the Brillouin zone edge is
approached. This means the peak in the tunneling will return to zero bias
when $B_{_{\Vert }}$ reaches a value such that $Q=\pm 2n\pi /a,$ with $%
n=0,1,2,3,\ldots $ provided $t_{N}\left( Q\right) $ is not zero. This is
illustrated in Fig.~3, which is produced using the first (normal) term in
Eq.~(\ref{oldt}) with a phenomenological broadening parameter $\delta $.
This non-monotonic behavior of the tunneling current peak is an unambiguous
signature of the ISP that distinquishes it from the IL. In this way one may
probe the transition from IL to ISP via measurement of the interlayer
tunneling current. Note that the tunneling current vanishes at large $%
B_{_{\Vert }}$ where $t_{N}\left( Q\right) \rightarrow 0$ \cite{jterm}.

The perturbative approach used in this section to compute the tunneling
current does not capture the full complexity of the phase diagram discussed
in Sec. V. To probe the various phase transitions discussed above and the
corresponding change in the collective excitations, the response functions
must be evaluated in a ground state that includes the parallel magnetic
field. That is, the response functions must be evaluated non-perturbatively
with respect to the tunneling matrix element $t_{0}$. Work on this problem
is currently in progress.

\section{Conclusion}

We have shown that, in higher Landau levels ($N>1$), a parallel magnetic
field can induce a complex sequence of phase transitions in the
two-dimensional electron gas of a double-quantum-well system. When the
separation between the wells or the parallel magnetic field are varied,
transitions occur between commensurate and incommensurate liquid or stripe
states. Working in the generalized random-phase approximation, we have
computed the spectrum of collective excitations of these four ground states
showing how the pseudospin-wave mode of the incommensurate liquid phase is
modified by the formation of the stripes in each well and by the parallel
magnetic field. We believe that the (incommensurate) stripe state could be
distinguished from the (incommensurate) liquid state in a tunneling
experiment because of the distinctive signature due to the periodicity of
the ISP.

\section{Acknowledgments}

This work was supported by a grant from the Natural Sciences and Engineering
Research Council of Canada (NSERC) and by NSF Grant No. DMR-0108451.

\appendix*

\section{Definition of the matrices $F$ and $D$}

With%
\begin{equation}
S_{\mathbf{q},\mathbf{q}^{\prime }}=\sin \left[ \frac{\mathbf{q}\times 
\mathbf{q}^{\prime }l_{\bot }^{2}}{2}\right]
\end{equation}%
The matrices $F$ and $D$ of Eq.~(\ref{simpler}) are given by

\begin{equation}
D_{i,1}\left( \mathbf{q,q}^{\prime }\right) =\hslash \left( 
\begin{array}{c}
2iS_{\mathbf{q},\mathbf{q}^{\prime }}\left\langle \widetilde{\rho }%
^{R,R}\left( \mathbf{q-q}^{\prime }\right) \right\rangle \\ 
-\left\langle \widetilde{\rho }^{R,L}\left( \mathbf{q-q}^{\prime }\right)
\right\rangle \gamma _{\mathbf{q},\mathbf{q}^{\prime }} \\ 
\left\langle \widetilde{\rho }^{L,R}\left( \mathbf{q-q}^{\prime }\right)
\right\rangle \gamma _{\mathbf{q},\mathbf{q}^{\prime }}^{\ast } \\ 
0%
\end{array}%
\right) ,
\end{equation}

\begin{equation}
D_{i,2}\left( \mathbf{q,q}^{\prime }\right) =\hslash \left( 
\begin{array}{c}
-\left\langle \widetilde{\rho }^{L,R}\left( \mathbf{q-q}^{\prime }\right)
\right\rangle \gamma _{\mathbf{q},\mathbf{q}^{\prime }} \\ 
\left\langle \widetilde{\rho }^{R,R}\left( \mathbf{q-q}^{\prime }\right)
\right\rangle \gamma _{\mathbf{q},\mathbf{q}^{\prime }}^{\ast }-\left\langle 
\widetilde{\rho }^{L,L}\left( \mathbf{q-q}^{\prime }\right) \right\rangle
\gamma _{\mathbf{q},\mathbf{q}^{\prime }} \\ 
0 \\ 
\left\langle \widetilde{\rho }^{L,R}\left( \mathbf{q-q}^{\prime }\right)
\right\rangle \gamma _{\mathbf{q},\mathbf{q}^{\prime }}^{\ast }%
\end{array}%
\right) ,
\end{equation}%
\begin{equation}
D_{i,3}\left( \mathbf{q,q}^{\prime }\right) =\hslash \left( 
\begin{array}{c}
\left\langle \widetilde{\rho }^{R,L}\left( \mathbf{q-q}^{\prime }\right)
\right\rangle \gamma _{\mathbf{q},\mathbf{q}^{\prime }}^{\ast } \\ 
0 \\ 
\left\langle \widetilde{\rho }^{L,L}\left( \mathbf{q-q}^{\prime }\right)
\right\rangle \gamma _{\mathbf{q},\mathbf{q}^{\prime }}^{\ast }-\left\langle 
\widetilde{\rho }^{R,R}\left( \mathbf{q-q}^{\prime }\right) \right\rangle
\gamma _{\mathbf{q},\mathbf{q}^{\prime }} \\ 
-\left\langle \widetilde{\rho }^{R,L}\left( \mathbf{q-q}^{\prime }\right)
\right\rangle \gamma _{\mathbf{q},\mathbf{q}^{\prime }}%
\end{array}%
\right) ,
\end{equation}%
\begin{equation}
D_{i,4}\left( \mathbf{q,q}^{\prime }\right) =\hslash \left( 
\begin{array}{c}
0 \\ 
\left\langle \widetilde{\rho }^{R,L}\left( \mathbf{q-q}^{\prime }\right)
\right\rangle \gamma _{\mathbf{q},\mathbf{q}^{\prime }}^{\ast } \\ 
-\left\langle \widetilde{\rho }^{L,R}\left( \mathbf{q-q}^{\prime }\right)
\right\rangle \gamma _{\mathbf{q},\mathbf{q}^{\prime }} \\ 
2iS_{\mathbf{q},\mathbf{q}^{\prime }}\left\langle \widetilde{\rho }%
_{m}^{L,L}\left( \mathbf{q-q}^{\prime }\right) \right\rangle%
\end{array}%
\right) ,
\end{equation}%
and (in these definitions, the tunneling $t_{N}\left( Q\right) $ and all
matrix elements are in units of $e^{2}/\kappa \ell $) 
\begin{eqnarray}
F_{1,1}\left( \mathbf{q},\mathbf{q}^{\prime }\right) &=&-2iS_{\mathbf{q},%
\mathbf{q}^{\prime }}\left[ \left[ H\left( \mathbf{q-q}^{\prime }\right)
-X\left( \mathbf{q-q}^{\prime }\right) -H\left( \mathbf{q}^{\prime }\right)
+X\left( \mathbf{q}^{\prime }\right) \right] \left\langle \widetilde{\rho }%
^{R,R}\left( \mathbf{q-q}^{\prime }\right) \right\rangle \right] \\
&&-2iS_{\mathbf{q},\mathbf{q}^{\prime }}\widetilde{H}\left( \mathbf{q-q}%
^{\prime }\right) \left( \gamma ^{\ast }\left( \mathbf{q-q}^{\prime }\right)
\right) ^{2}\left\langle \widetilde{\rho }^{L,L}\left( \mathbf{q-q}^{\prime
}\right) \right\rangle ,  \notag
\end{eqnarray}%
\begin{eqnarray}
F_{2,1}\left( \mathbf{q},\mathbf{q}^{\prime }\right) &=&-t_{N}\left(
Q\right) \delta _{\mathbf{q},\mathbf{q}^{\prime }}-\left[ H\left( \mathbf{q}%
^{\prime }\right) -X\left( \mathbf{q}^{\prime }\right) +\widetilde{X}%
_{-}\left( \mathbf{q-q}^{\prime }\right) \right] \left\langle \widetilde{%
\rho }^{R,L}\left( \mathbf{q-q}^{\prime }\right) \right\rangle \gamma _{%
\mathbf{q},\mathbf{q}^{\prime }}  \notag \\
&&+\left[ \widetilde{H}\left( \mathbf{q}^{\prime }\right) \gamma ^{2}\left( 
\mathbf{q}^{\prime }\right) \right] \left\langle \widetilde{\rho }%
^{R,L}\left( \mathbf{q-q}^{\prime }\right) \right\rangle \gamma _{\mathbf{q},%
\mathbf{q}^{\prime }}^{\ast },
\end{eqnarray}%
\begin{eqnarray}
F_{3,1}\left( \mathbf{q},\mathbf{q}^{\prime }\right) &=&t_{N}\left( Q\right)
\delta _{\mathbf{q},\mathbf{q}^{\prime }}+\left[ H\left( \mathbf{q}^{\prime
}\right) -X\left( \mathbf{q}^{\prime }\right) +\widetilde{X}_{+}\left( 
\mathbf{q-q}^{\prime }\right) \right] \left\langle \widetilde{\rho }%
^{L,R}\left( \mathbf{q-q}^{\prime }\right) \right\rangle \gamma _{\mathbf{q},%
\mathbf{q}^{\prime },}^{\ast }  \notag \\
&&-\left[ \widetilde{H}\left( \mathbf{q}^{\prime }\right) \gamma ^{2}\left( 
\mathbf{q}^{\prime }\right) \right] \left\langle \widetilde{\rho }%
^{L,R}\left( \mathbf{q-q}^{\prime }\right) \right\rangle \gamma _{\mathbf{q},%
\mathbf{q}^{\prime }},
\end{eqnarray}%
\begin{equation}
F_{4,1}\left( \mathbf{q},\mathbf{q}^{\prime }\right) =2iS_{\mathbf{q},%
\mathbf{q}^{\prime }}\left[ \widetilde{H}\left( \mathbf{q}^{\prime }\right)
\gamma ^{2}\left( \mathbf{q}^{\prime }\right) \right] \left\langle 
\widetilde{\rho }^{L,L}\left( \mathbf{q-q}^{\prime }\right) \right\rangle ,
\end{equation}%
\begin{equation}
F_{1,2}\left( \mathbf{q},\mathbf{q}^{\prime }\right) =-t_{N}\left( Q\right)
\delta _{\mathbf{q},\mathbf{q}^{\prime }}-\left[ \widetilde{X}_{+}\left( 
\mathbf{q-q}^{\prime }\right) -\widetilde{X}_{-}\left( \mathbf{q}^{\prime
}\right) \right] \left\langle \widetilde{\rho }^{L,R}\left( \mathbf{q-q}%
^{\prime }\right) \right\rangle \gamma _{\mathbf{q},\mathbf{q}^{\prime }},
\end{equation}%
\begin{eqnarray}
F_{2,2}\left( \mathbf{q},\mathbf{q}^{\prime }\right) &=&-\left[ H\left( 
\mathbf{q-q}^{\prime }\right) -X\left( \mathbf{q-q}^{\prime }\right) +%
\widetilde{X}_{-}\left( \mathbf{q}^{\prime }\right) \right] \left\langle 
\widetilde{\rho }^{R,R}\left( \mathbf{q-q}^{\prime }\right) \right\rangle
\gamma _{\mathbf{q},\mathbf{q}^{\prime }}^{\ast } \\
&&+\widetilde{H}\left( \mathbf{q-q}^{\prime }\right) \left( \gamma \left( 
\mathbf{q-q}^{\prime }\right) \right) ^{2}\left\langle \widetilde{\rho }%
^{R,R}\left( \mathbf{q-q}^{\prime }\right) \right\rangle \gamma _{\mathbf{q},%
\mathbf{q}^{\prime }}  \notag \\
&&+\left[ H\left( \mathbf{q-q}^{\prime }\right) -X\left( \mathbf{q-q}%
^{\prime }\right) +\widetilde{X}_{-}\left( \mathbf{q}^{\prime }\right) %
\right] \left\langle \widetilde{\rho }^{L,L}\left( \mathbf{q-q}^{\prime
}\right) \right\rangle \gamma _{\mathbf{q},\mathbf{q}^{\prime }}  \notag \\
&&-\widetilde{H}\left( \mathbf{q-q}^{\prime }\right) \left( \gamma ^{\ast
}\left( \mathbf{q-q}^{\prime }\right) \right) ^{2}\left\langle \widetilde{%
\rho }^{L,L}\left( \mathbf{q-q}^{\prime }\right) \right\rangle \gamma _{%
\mathbf{q},\mathbf{q}^{\prime }}^{\ast },  \notag
\end{eqnarray}

\begin{equation}
F_{3,2}\left( \mathbf{q},\mathbf{q}^{\prime }\right) =0,
\end{equation}%
\begin{equation}
F_{4,2}\left( \mathbf{q},\mathbf{q}^{\prime }\right) =t_{N}\left( Q\right)
\delta _{\mathbf{q},\mathbf{q}^{\prime }}+\left( \frac{e^{2}}{\kappa l_{\bot
}}\right) \left[ \widetilde{X}_{+}\left( \mathbf{q-q}^{\prime }\right) -%
\widetilde{X}_{-}\left( \mathbf{q}^{\prime }\right) \right] \left\langle 
\widetilde{\rho }^{L,R}\left( \mathbf{q-q}^{\prime }\right) \right\rangle
\gamma _{\mathbf{q},\mathbf{q}^{\prime }}^{\ast },
\end{equation}%
\begin{equation}
F_{1,3}\left( \mathbf{q},\mathbf{q}^{\prime }\right) =t_{N}\left( Q\right)
\delta _{\mathbf{q},\mathbf{q}^{\prime }}+\left( \frac{e^{2}}{\kappa l_{\bot
}}\right) \left[ \widetilde{X}_{-}\left( \mathbf{q-q}^{\prime }\right) -%
\widetilde{X}_{+}\left( \mathbf{q}^{\prime }\right) \right] \left\langle 
\widetilde{\rho }^{R,L}\left( \mathbf{q-q}^{\prime }\right) \right\rangle
\gamma _{\mathbf{q},\mathbf{q}^{\prime }}^{\ast },
\end{equation}%
\begin{equation}
F_{2,3}\left( \mathbf{q},\mathbf{q}^{\prime }\right) =0,
\end{equation}%
\begin{eqnarray}
F_{3,3}\left( \mathbf{q},\mathbf{q}^{\prime }\right) &=&-\left[ H\left( 
\mathbf{q-q}^{\prime }\right) -X\left( \mathbf{q-q}^{\prime }\right) +%
\widetilde{X}_{+}\left( \mathbf{q}^{\prime }\right) \right] \left\langle 
\widetilde{\rho }^{L,L}\left( \mathbf{q-q}^{\prime }\right) \right\rangle
\gamma _{\mathbf{q},\mathbf{q}^{\prime }}^{\ast } \\
&&+\widetilde{H}\left( \mathbf{q-q}^{\prime }\right) \left( \gamma ^{\ast
}\left( \mathbf{q-q}^{\prime }\right) \right) ^{2}\left\langle \widetilde{%
\rho }^{L,L}\left( \mathbf{q-q}^{\prime }\right) \right\rangle \gamma _{%
\mathbf{q},\mathbf{q}^{\prime }}  \notag \\
&&+\left[ H\left( \mathbf{q-q}^{\prime }\right) -X\left( \mathbf{q-q}%
^{\prime }\right) +\widetilde{X}_{+}\left( \mathbf{q}^{\prime }\right) %
\right] \left\langle \widetilde{\rho }^{R,R}\left( \mathbf{q-q}^{\prime
}\right) \right\rangle \gamma _{\mathbf{q},\mathbf{q}^{\prime }}  \notag \\
&&-\widetilde{H}\left( \mathbf{q-q}^{\prime }\right) \left( \gamma \left( 
\mathbf{q-q}^{\prime }\right) \right) ^{2}\left\langle \widetilde{\rho }%
^{R,R}\left( \mathbf{q-q}^{\prime }\right) \right\rangle \gamma _{\mathbf{q},%
\mathbf{q}^{\prime }}^{\ast },  \notag
\end{eqnarray}%
\qquad 
\begin{equation}
F_{4,3}\left( \mathbf{q},\mathbf{q}^{\prime }\right) =-t_{N}\left( Q\right)
\delta _{\mathbf{q},\mathbf{q}^{\prime }}-\left[ \widetilde{X}_{-}\left( 
\mathbf{q-q}^{\prime }\right) -\widetilde{X}_{+}\left( \mathbf{q}^{\prime
}\right) \right] \left\langle \widetilde{\rho }^{R,L}\left( \mathbf{q-q}%
^{\prime }\right) \right\rangle \gamma _{\mathbf{q},\mathbf{q}^{\prime }},
\end{equation}%
\begin{equation}
F_{1,4}\left( \mathbf{q},\mathbf{q}^{\prime }\right) =2iS_{\mathbf{q},%
\mathbf{q}^{\prime }}\left\langle \widetilde{\rho }^{R,R}\left( \mathbf{q-q}%
^{\prime }\right) \right\rangle \left[ \widetilde{H}\left( \mathbf{q}%
^{\prime }\right) \left( \gamma ^{\ast }\left( \mathbf{q}^{\prime }\right)
\right) ^{2}\right] ,
\end{equation}%
\begin{eqnarray}
F_{2,4}\left( \mathbf{q},\mathbf{q}^{\prime }\right) &=&t_{N}\left( Q\right)
\delta _{\mathbf{q},\mathbf{q}^{\prime }}+\left[ \widetilde{X}_{-}\left( 
\mathbf{q-q}^{\prime }\right) +H\left( \mathbf{q}^{\prime }\right) -X\left( 
\mathbf{q}^{\prime }\right) \right] \left\langle \widetilde{\rho }%
^{R,L}\left( \mathbf{q-q}^{\prime }\right) \right\rangle \gamma _{\mathbf{q},%
\mathbf{q}^{\prime }}^{\ast },  \notag \\
&&-\left[ \widetilde{H}\left( \mathbf{q}^{\prime }\right) \left( \gamma
^{\ast }\left( \mathbf{q}^{\prime }\right) \right) ^{2}\right] \left\langle 
\widetilde{\rho }^{R,L}\left( \mathbf{q-q}^{\prime }\right) \right\rangle
\gamma _{\mathbf{q},\mathbf{q}^{\prime }},
\end{eqnarray}%
\begin{eqnarray}
F_{3,4}\left( \mathbf{q},\mathbf{q}^{\prime }\right) &=&-t_{N}\left(
Q\right) \delta _{\mathbf{q},\mathbf{q}^{\prime }}-\left[ \widetilde{X}%
_{+}\left( \mathbf{q-q}^{\prime }\right) +H\left( \mathbf{q}^{\prime
}\right) -X\left( \mathbf{q}^{\prime }\right) \right] \left\langle 
\widetilde{\rho }^{L,R}\left( \mathbf{q-q}^{\prime }\right) \right\rangle
\gamma _{\mathbf{q},\mathbf{q}^{\prime }},  \notag \\
&&+\left[ \widetilde{H}\left( \mathbf{q}^{\prime }\right) \left( \gamma
^{\ast }\left( \mathbf{q}^{\prime }\right) \right) ^{2}\right] \left\langle 
\widetilde{\rho }^{L,R}\left( \mathbf{q-q}^{\prime }\right) \right\rangle
\gamma _{\mathbf{q},\mathbf{q}^{\prime }}^{\ast }
\end{eqnarray}%
\begin{eqnarray}
F_{4,4}\left( \mathbf{q},\mathbf{q}^{\prime }\right) &=&-2iS_{\mathbf{q},%
\mathbf{q}^{\prime }}\widetilde{H}\left( \mathbf{q-q}^{\prime }\right)
\gamma ^{2}\left( \mathbf{q-q}^{\prime }\right) \left\langle \widetilde{\rho 
}^{R,R}\left( \mathbf{q-q}^{\prime }\right) \right\rangle \\
&&-2iS_{\mathbf{q},\mathbf{q}^{\prime }}\left[ H\left( \mathbf{q-q}^{\prime
}\right) -X\left( \mathbf{q-q}^{\prime }\right) -H\left( \mathbf{q}^{\prime
}\right) +X\left( \mathbf{q}^{\prime }\right) \right] \left\langle 
\widetilde{\rho }^{L,L}\left( \mathbf{q-q}^{\prime }\right) \right\rangle . 
\notag
\end{eqnarray}


\begin{thebibliography}{99}
\bibitem{com} Further evidence of this is found in the static HF solutions,
which show a tendency to develop weak oscillations in the in-plane
components of the pseudospin just before the transition to the stripe phase,
which we interpret as evidence of the SLS instability. These oscillations
have only a very small effect on the energy.
\bibitem{comlw} For $t_N(\mathbf{Q})=0$, $\omega _{CL}\left( \mathbf{q}%
\right)$ is gapless. But since the CL is always unstable to the IL in this
limit, such gaplessness is never observable in the CL state.
\bibitem{jterm} We remark that in the ISP, $\chi _{R,L,R,L}\left( \mathbf{q},%
\mathbf{q}^{\prime },\omega \right) =\chi _{R,L,R,L}\left( \mathbf{k+G},%
\mathbf{k+G}^{\prime },\omega \right) $ where $\mathbf{k}$ is a vector in
the first Brillouin zone in the reciprocal lattice and $\mathbf{G,G}^{\prime
}$ are reciprocal lattice vectors. In computing the tunneling current in
Fig.~3, we ignored the second term in Eq.~(\ref{oldt}) that can contribute
to the current only at some discrete values of $Q$.
\end{thebibliography}
\end{document}